\documentclass[twocolumn,mathlines]{aastex631}

\usepackage{amsmath,epsfig,url}
\usepackage{float}
\usepackage{graphicx,epstopdf,float,color}
\usepackage{rotating,multirow}
\usepackage{verbatim}

\usepackage{hyperref}

\usepackage[acronym,nomain,nopostdot]{glossaries}
\glsdisablehyper
\newacronym{lr}{LR}{Low-Resolution}
\newacronym{hr}{HR}{High-Resolution}
\newacronym{sr}{SR}{Super-Resolution}
\newacronym{jwst}{JWST}{James Webb Space Telescope}
\newacronym{jades}{JADES}{JWST Advanced Deep Extragalactic Survey}
\newacronym{gto}{GTO}{Guaranteed Time Observations}
\newacronym{goods}{GOODS}{Great Observatories Origins Deep Survey}
\newacronym{sdss}{SDSS}{Sloan Digital Sky Survey}
\newacronym{desi}{DESI}{Dark Energy Spectroscopic Instrument}
\newacronym{vlt}{VLT}{Very Large Telescope}
\newacronym{agn}{AGN}{Active Galactic Nucleus}
\newacronym{bpt}{BPT}{Baldwin--Phillips--Terlevich}
\newacronym{cnn}{CNN}{Convolutional Neural Network}
\newacronym{gpu}{GPU}{Graphics Processing Unit}
\newacronym{psd}{PSD}{Power Spectral Density}
\newacronym{nmad}{NMAD}{Normalized Median Absolute Deviation}
\newacronym{uv}{UV}{Ultraviolet}
\newacronym{sr2}{SR2}{Super-Resolution 2}
\newacronym{sr1}{SR1}{Super-Resolution 1}
\newacronym{dr}{DR}{Data Release}
\newacronym{candels}{CANDELS}{Cosmic Assembly Near-infrared Deep Extragalactic Legacy Survey}
\newacronym{nirspec}{NIRSpec}{Near-Infrared Spectrograph}
\newacronym{nircam}{NIRCam}{Near-Infrared Camera}
\newacronym{fwhm}{FWHM}{Full Width at Half Maximum}

\shorttitle{Super-Resolving Galaxy Spectra}
\shortauthors{Haghjoo et al.}

\begin{document}


\title{Learning to See Sharper:\\
A Physics-Informed Artificial Intelligence Framework for Super-Resolving Galaxy Spectra}
\author[0009-0006-3071-7143]{Aryana Haghjoo}
\affiliation{Department of Physics and Astronomy, University of California Riverside, Riverside, CA 92521, USA}

\author[0000-0003-2226-5395]{Shoubaneh Hemmati}
\affiliation{IPAC, California Institute of Technology, Pasadena, CA 91125, USA}

\author[0000-0001-5846-4404]{Bahram Mobasher}
\affiliation{Department of Physics and Astronomy, University of California Riverside, Riverside, CA 92521, USA}

\author[0000-0003-3691-937X]{Nima Chartab}
\affiliation{IPAC, California Institute of Technology, Pasadena, CA 91125, USA}

\author[0000-0002-1894-3301]{Tim Eifler}
\affiliation{The University of Arizona, Tucson, AZ 85721, USA}

\author[0000-0003-0447-8426]{Emily Ramey}
\affiliation{Department of Astronomy, University of California Berkeley, Berkeley, CA 94720, USA}

\author[0000-0001-8242-9983]{Hooshang Nayyeri}
\affiliation{Amazon}

\author[0000-0002-0364-1159]{Zahra Sattari}
\affiliation{IPAC, California Institute of Technology, Pasadena, CA 91125, USA}

\email{aryana.haghjoo@email.ucr.edu}

\journalinfo{\textcopyright\ 2026. All rights reserved. To be submitted to the Astrophysical Journal.}

\begin{abstract}
The information recoverable from galaxy spectra depends fundamentally on spectral resolution, yet assembling large samples at high resolution remains observationally expensive. We present a deep-learning framework for spectral super-resolution that enhances low-resolution galaxy spectra by a factor of $\sim$10 in resolving power ($R\sim100$ to $R\sim1000$). The model is trained on 2,858 paired \gls{jwst}/\gls{nirspec} observations from the \gls{jades} program, where low-resolution prism spectra are matched with medium-resolution grating spectra (G140M, G235M, G395M) combined into a unified reference covering 1.0--5.3\,$\mu$m. Our three-stage architecture performs an initial super-resolution, infers the redshift from the coarse reconstruction, and then applies a physics-informed residual refinement that uses attention across emission-line tokens to learn inter-line relationships and predict parametric line profiles, alongside a convolutional branch for continuum corrections. Evaluated on a 20\% held-out sample, the model achieves noise-limited residuals over most of the spectral range and improves the signal-to-noise ratio of key diagnostic lines including [O\,\textsc{ii}], H$\beta$, [O\,\textsc{iii}], and H$\alpha$. The super-resolved spectra separate features that are entirely unresolved at prism resolution, such as the [O\,\textsc{iii}]\,$\lambda\lambda4959,5007$ doublet and H$\beta$, in exchange for a modest retrieval of line fluxes. As a proof of concept using \gls{jwst} data, this approach is readily extensible to the low-resolution grism spectroscopy that will be delivered by \textit{Euclid} and the \textit{Roman Space Telescope}, potentially enabling population-level diagnostics across millions of galaxy spectra that would otherwise be inaccessible at low-resolution.
\end{abstract}

\glsresetall

\keywords{galaxies:, methods: data analysis, methods: statistical}
\section{Introduction}\label{intro}

The information content recoverable from galaxy light is fundamentally set by the resolution at which it is observed. From broadband photometry through narrow-band imaging to spectroscopy, progressively finer spectral features become accessible, and with them a more informative set of physical diagnostics. Emission lines detected in galaxy spectra provide a wealth of information: precise redshifts for cosmological analyses such as galaxy clustering and weak lensing (e.g., \citealt{Hemmati2019,Wang2019,DESI2025}), and measurements of physical properties including dust attenuation from the Balmer decrement (e.g., \citealt{Shivaei2016}; \citealt{Alavi2026}), star-formation rate (e.g., \citealt{Khostovan2024,Woodrum2025,Shen2025}), gas-phase metallicity (e.g., \citealt{PettiniPagel2004,Tremonti2004}; \citealt{Shivaei2020}), \gls{agn} activity via \gls{bpt} diagnostics (e.g., \citealt{BPT_original,agn_classification,agn_misclass}), kinematics (e.g., \citealt{kinematics_1, kinematics_2}), and electron density from [S\,\textsc{ii}] line ratios (e.g., \citealt{electron_density_sii, Darvish2015}). However, many of these measurements depend on well-resolved lines that are close in wavelength, such as H$\alpha$ and [N\,\textsc{ii}] (e.g., \citealt{Sobral2009,faisst2018empirical,HN_blending}), or faint features such as the [O\,\textsc{i}] shock tracer associated with stellar winds and outflows (e.g., \citealt{shochandfeedback_2,shochandfeedback_3}; \citealt{shochandfeedback_1}). When the spectral resolution is insufficient, these lines blend together, propagating systematic biases into every derived quantity, from the redshifts that underpin cosmological constraints to the dust corrections and star-formation rates that govern our understanding of galaxy evolution. Achieving the full scientific return of spectroscopic surveys therefore demands high spectral resolution. Equally important are large samples that capture the diversity of galaxy populations and distinguish genuine physical correlations from stochastic variations.

Assembling large spectroscopic samples that simultaneously achieve high spectral resolution, wide area coverage, and sufficient depth remains one of the key observational challenges in extragalactic astronomy. Each current and forthcoming facility navigates a different compromise among these requirements. Ground-based surveys such as \gls{sdss} (\citealt{SDSS}) and \gls{desi} (\citealt{DESI}) cover wide areas at moderate-to-high spectral resolution, but are limited in depth and therefore probe primarily the low-redshift universe. Space-based survey missions such as \textit{Euclid} (\citealt{euclid}) and the \textit{Roman Space Telescope} (\citealt{roman_1, roman_2}) reach fainter galaxies over vast areas, but rely on slitless grism spectroscopy that provides only low spectral resolution, trading resolving power for survey volume. \textit{\glsentrylong{jwst}} (\glsentryshort{jwst}; \citealt{jwst})\glsunset{jwst}, by contrast, offers high spectral resolution, but its limited field of view restricts it to targeted observations rather than wide-area surveys. MOONS at the \gls{vlt} (\citealt{moons}) and 4MOST (\citealt{fourmost}) will soon assemble spectroscopic samples orders of magnitude larger at comparable resolving power, but their coverage is bounded: MOONS spans 0.64--1.8\,$\mu$m and 4MOST is optical, so the rest-frame optical diagnostics above leave range at $z\gtrsim1.7$ and $z\gtrsim0.4$ respectively. What remains inaccessible is therefore not large spectroscopic samples in general, but large high-resolution samples at high redshift, where wide-area slitless grism spectroscopy is the only practical option. As a result, no single facility currently delivers high-resolution spectra for statistically large samples at high redshift, and therefore, cosmological probes such as galaxy clustering and weak lensing must contend with either blended emission lines or limited sample sizes.

One way to circumvent this trade-off is to leverage the small but growing number of galaxies observed at both low and high spectral resolution, and use machine learning to learn the mapping between the two, enabling the  resolution enhancement of the much larger low-resolution samples. Machine-learning methods have already demonstrated success across a range of astronomical applications, from imaging super-resolution and source deblending to galaxy classification and the estimation of physical parameters (e.g., \citealt{applicationsAIinAstro, Hemmati2022, sogol_1, astroclip, highly_related, ion1, shooby_agn}). Spectral super-resolution is an established class of methods outside astronomy, where it usually denotes recovering a densely sampled spectrum from a few broadband channels, most often hyperspectral data from ordinary RGB images (e.g., \citealt{arad2016sparse}; see \citealt{spectral_reconstruction_survey} for a review). What is novel here is the astrophysical setting: the features to be recovered are emission lines whose observed positions depend on a redshift that is not known in advance. In this work, we develop a deep-learning-based spectral super-resolution framework that enhances low-resolution near-infrared galaxy spectra to medium spectral resolution, effectively reconstructing important diagnostic features that would otherwise be blended.

Super-resolution is an ill-posed problem. Many different high-resolution spectra are consistent with the same low-resolution input. A pixel-wise loss therefore returns roughly their average, which smooths away the sharp features the method is meant to recover (\citealt{ledig2017srgan}). Losses that instead reward realistic-looking output trade accuracy for realism (\citealt{blau2018perception}) and can invent features that are not in the data (\citealt{cohen2018hallucinate}). In spectroscopy such a hallucinated line is the more damaging failure, because it propagates into every quantity derived from it. Our refinement stage therefore models each emission line explicitly, with a gate that can decline to place the line at all (\S\ref{sec:architecture}).

Neural networks have also been used to infer redshifts directly from spectra (\citealt{zestimatr}). Examples include a multi-task network that returns a redshift probability distribution together with physical parameters (\citealt{Ginolfi2025}), and a pre-trained network that reconstructs spectra, estimates redshifts and flags anomalies (\citealt{Zhong2025}).

The model is trained on paired spectra from the \textit{\glsentrylong{jades}} (\glsentryshort{jades}; \citealt{jades_overview})\glsunset{jades}, where low-resolution prism observations ($R\sim100$) are matched with medium-resolution grating spectra (G140M, G235M, G395M; $R\sim1000$) of the same galaxies. By learning from these paired samples, augmented with noise and redshift perturbations, the network captures physically meaningful mappings between low- and higher-resolution spectral representations. In this proof of concept study, we use \gls{jwst} prism and grating data, but the approach can further be extended to the low-resolution grism spectroscopy delivered by wide-area cosmological surveys such as \textit{Euclid} and the \textit{Roman Space Telescope}, potentially enabling population-level diagnostics that would otherwise be inaccessible at grism resolution.

The paper is organized as follows. \S\ref{sec:data} describes the \gls{jades} spectral data used in this study. \S\ref{sec:method} outlines our methodology, including data preprocessing, augmentation, and the model architecture and training procedure. \S\ref{sec:results} presents results and evaluation of our super-resolution model. Finally, \S\ref{sec:discussions} discusses the implications of this work and avenues for future improvement. The framework is released as an open-source Python package, \texttt{specsr}, which reproduces every result and figure presented here (\S\ref{sec:software}).
\section{The JADES Dataset}\label{sec:data}

\begin{figure*}[ht!]
    \centering
    \includegraphics[width=\linewidth]{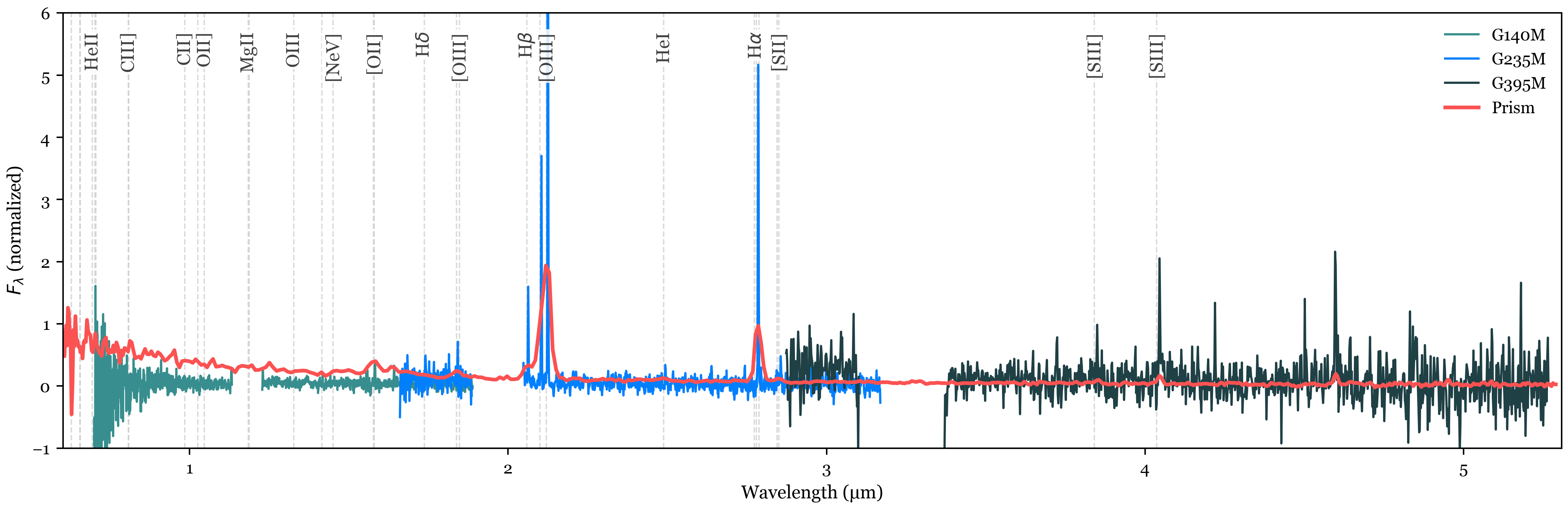}
    \caption{Example low- and medium-resolution \gls{jwst} \gls{nirspec} spectra for a representative \gls{jades} galaxy. The prism spectrum is shown in red, and the medium-resolution G140M, G235M, and G395M segments are over-plotted in green, blue, and black, respectively. 
    The natural gaps and overlaps between the three medium-resolution gratings are evident across the wavelength range. Vertical dashed lines mark key emission lines commonly utilized in nebular and galaxy evolution studies, shifted to the observed frame at the galaxy's redshift ($z=3.235$). All spectra are normalized to allow visual comparison.}
    \label{fig:matched_spectra}
\end{figure*}

\begin{figure*}[ht!]
    \centering
    \includegraphics[width=\linewidth]{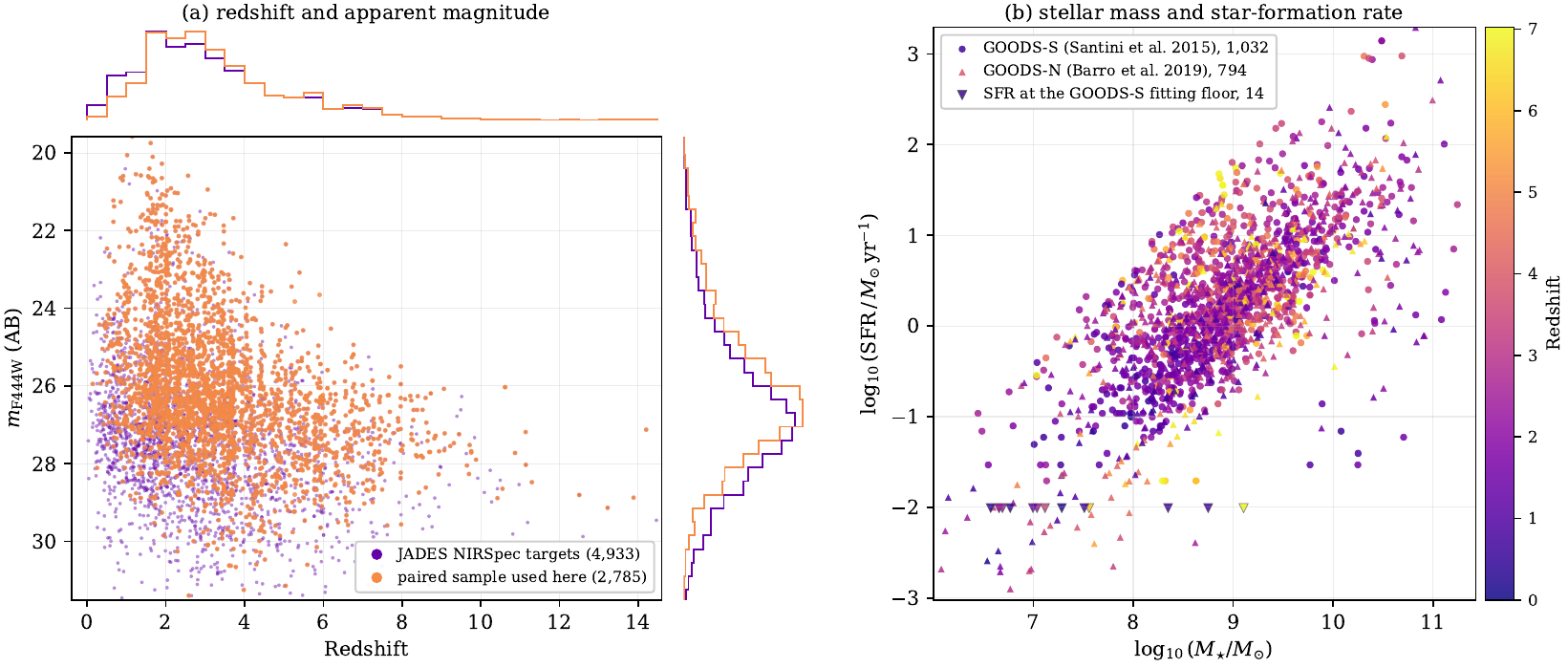}
    \caption{The galaxies this work uses, set against the parent \gls{jades} \gls{nirspec} sample.
    \emph{(a)} Redshift against apparent F444W magnitude, with marginal distributions above and to the right. Gray points are all \gls{jades} \gls{nirspec} targets; colored points are the $2{,}858$ galaxies observed in the prism and all three medium-resolution gratings that this work uses. Magnitudes are Kron fluxes from the \gls{jades} \gls{dr}5 \gls{nircam} catalogs and are available for $97.4\%$ of the sample.
    \emph{(b)} Stellar mass against star-formation rate, colored by redshift, for the $64\%$ of the sample with a \gls{candels} (\citealt{Grogin2011, Koekemoer2011}) counterpart; eight galaxies with extreme fitted values fall outside the plotted axes. The two fields are marked separately because their values come from different spectral-energy-distribution fits: FAST for \gls{goods}-North (\citealt{Barro2019}) and the median-of-methods masses with the $6a\_\tau$ star-formation rates for \gls{goods}-South (\citealt{Santini2015}). The two symbol sets should therefore not be read as one measurement scale. Downward triangles mark galaxies sitting on the \gls{goods}-South fitting floor of $0.01\,M_{\odot}\,\mathrm{yr}^{-1}$, which are upper limits, not measurements. The matched fraction falls from $78\%$ below $z=2$ to $33\%$ above $z=6$, so this panel characterizes the lower-redshift part of the sample better than the higher-redshift part.}
    \label{fig:sample}
\end{figure*}

For a comprehensive description of the \gls{jades}, we refer the reader to \citealt{jades_overview, jades_nircam, jades_nirspec, jades_nircam_2, jades_nirspec_2, Curtis-Lake2025}. Briefly, \gls{jades} provides deep imaging and spectroscopy in the \gls{goods}-North and \gls{goods}-South fields (\citealt{GOODS}), combining \gls{nircam} and \gls{nirspec} observations that span from rest-frame ultraviolet to near-infrared. The complete \gls{jades} spectroscopic sample (Data Release 4; \citealt{Curtis-Lake2025}) consists of 5,190 targets, of which 3,297 have robust spectroscopic redshifts extending to $z \sim 14$.

What makes the \gls{jades} particularly suited to this work is the tiered survey design that pairs low-resolution prism observations ($R\sim30$--$300$) with medium-resolution grating modes ($R\sim1000$; G140M, G235M, and G395M) for the same targets. The survey architecture balances depth and area: Deep and Ultra-Deep tiers comprise $\sim$700 galaxies with 20--50 hour integrations in the prism mode, while Medium tiers include over 4,400 galaxies with shallower 1--3 hour exposures. This diversity ensures the training set covers a wide dynamic range in signal-to-noise. An example of the prism and medium-resolution spectra for a representative \gls{jades} galaxy is shown in Figure~\ref{fig:matched_spectra}.

Figure~\ref{fig:sample} characterizes the galaxies this work uses. Requiring the prism and all three medium-resolution gratings for the same target leaves $2{,}858$ galaxies of the $5{,}190$ \gls{jades} \gls{nirspec} targets, spanning $z=0.14$--$14.18$ with a median of $2.95$ and a median F444W magnitude of $26.3$ (5th--95th percentiles $22.7$--$28.7$). Panel~(a) shows that this requirement does not select a distinct population out of the parent sample: the two redshift distributions track one another, and the paired sample is fainter than $m_{\rm F444W}=28$ in its lowest decile. Where a \gls{candels} counterpart exists, the galaxies occupy the star-forming main sequence, with a median $\log_{10}(M_{\star}/M_{\odot})$ of $8.9$ and a median $\log_{10}({\rm SFR}/M_{\odot}\,{\rm yr}^{-1})$ of $0.24$. That is the selection to keep in mind when reading \S\ref{sec:results}: \gls{jades} targeting favors actively star-forming systems, and the paired sample inherits that preference, so the performance reported here is measured on emission-line galaxies, not quiescent or absorption-dominated systems.

The prism observations capture the full 0.6--5.3\,$\mu$m wavelength range, while the medium-resolution gratings extend coverage to 5.5\,$\mu$m, together providing access to key diagnostic lines such as H$\alpha$, H$\beta$, [O\,\textsc{iii}], [N\,\textsc{ii}], and [S\,\textsc{ii}]. The low resolving power of the prism results in the blending of critical features, including H$\alpha$--[N\,\textsc{ii}], [O\,\textsc{iii}], and the [O\,\textsc{ii}] and [S\,\textsc{ii}] doublets. In contrast, the medium-resolution gratings resolve these lines individually. By combining the three grating modes, we construct a single medium-resolution reference spectrum covering 1.0--5.3\,$\mu$m, which serves as the high-resolution target for model training. The three arms do not tile this range perfectly for every galaxy; the stitching procedure and the treatment of the wavelengths that remain uncovered are described in \S\ref{sec:preproc}. As noted in \S\ref{intro}, the broad wavelength coverage of the \gls{jades} training data makes this framework readily extendable to low-resolution grism spectroscopy from missions such as the \textit{Euclid} and \textit{Roman Space Telescope}.
\section{Methodology}\label{sec:method}
Our spectral super-resolution framework is described in three parts: \emph{preprocessing} (\S\ref{sec:preproc}), \emph{augmentation} (\S\ref{sec:augmentation}), and \emph{model architecture and training} (\S\ref{sec:architecture}). The hyperparameter search that fine tunes the network's free parameters is a supporting study that is reported in Appendix~\ref{sec:optimization}.
\subsection{Pre-processing}\label{sec:preproc}
The \gls{jades} dataset provides low-resolution \textit{prism} spectra and three medium-resolution grating spectra for an overlapping subset of galaxies. To construct training pairs suitable for spectral super-resolution, we preprocess the data through source matching, medium-resolution spectrum construction, wavelength alignment, and quality filtering. The goal of this pipeline is to produce wavelength-aligned low- and medium-resolution spectra that are directly comparable.

We first identify galaxies observed in both \textit{prism} mode and all three medium-resolution gratings. Exposures are grouped by the pair (field, target identifier), so that a galaxy observed under more than one survey tier is counted once; $33$ \gls{jades} \gls{dr}4 targets are observed in two tiers. For each galaxy we adopt the catalog spectroscopic redshift $z_{\rm Spec}$ with quality flag A, B or C, the same definition behind the $3{,}297$ robust redshifts quoted in \S\ref{sec:data}.

For every matched object, the three grating spectra (G140M, G235M, G395M) are merged into a single reference by \textit{inverse-variance weighting} where they overlap, so a noisy arm edge is down-weighted automatically. The stitched reference covers a mean $96.9\%$ of the grid; the remaining wavelengths, which no detector measured for that galaxy, are carried as an explicit validity mask and \emph{excluded from the loss} instead of being filled.

Both spectra are then placed on a common \textit{logarithmic} wavelength grid, with constant $\lambda/\Delta\lambda = 4000$ and $6{,}671$ samples spanning $1.0$--$5.3$\,$\mu$m. The sampling is fine enough that the $R \sim 1000$ reference is never down-sampled; only the prism spectrum is up-sampled onto this grid. Resampling is \textit{flux-conserving}, integrating the input over the overlap of wavelength bins, because the physically meaningful quantity in an emission line is its integrated flux: on native G140M sampling this grid loses $0.03\%$ of the integrated flux of an $R \sim 1000$ line and retains $97.8\%$ of its peak.

The grid starts at $1.0$\,$\mu$m although the prism reaches $0.60$\,$\mu$m, because the prism is noise-dominated below that: its median per-sample $|f_\lambda|/\sigma_\lambda$ is $0.93$ over $0.60$--$0.70$\,$\mu$m against $2.50$ over $1.00$--$1.05$\,$\mu$m. The bound has a cost, since extending it to $0.6$\,$\mu$m would bring [O\,\textsc{ii}]\,$\lambda3727$ into range for a further $14.4\%$ of our galaxies. The upper limit of $5.3$\,$\mu$m is the prism's nominal limit.

Finally, we require that a galaxy be measured over at least half of the grid in both the stitched reference and the resampled prism spectrum. No emission-line significance cut is applied.

For each galaxy that passes all selection criteria, the pipeline outputs a standardized data product consisting of the up-sampled prism flux and uncertainty on the unified wavelength grid, the stitched medium-resolution flux and uncertainty on the same grid, and associated metadata including sky coordinates, redshift, and unique identifiers. These wavelength-aligned, quality-filtered spectral pairs form the final dataset used for augmentation and model training in the following sections.

\subsection{Augmentation}\label{sec:augmentation}

\begin{figure*}[ht!]
    \centering
    \includegraphics[width=\linewidth]{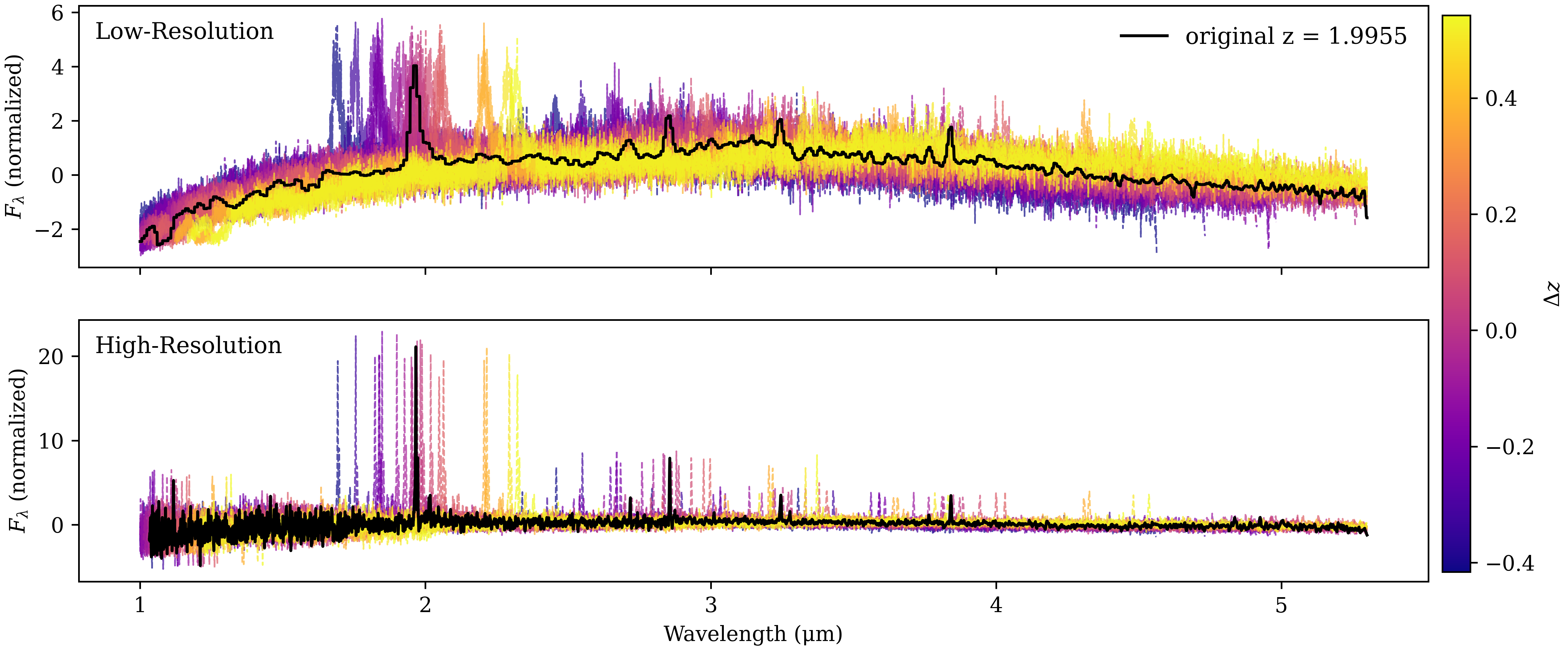}
    \caption{Example of the data augmentation procedure applied to a matched pair of low- and high-resolution \gls{jades} spectra. \emph{Top}: the original low-resolution prism spectrum (solid black) and twenty augmented realizations (colored dashed curves) generated by applying Gaussian-distributed redshift offsets ($z_{\mathrm{shift}}\sim\mathcal{N}(0,\sigma_z)$ with $\sigma_z=0.3$) and flux perturbations scaled to 10\% of the local flux. \emph{Bottom}: the corresponding high-resolution spectrum. Each realization applies the \emph{same} redshift offset to both members of the pair; the flux perturbations are drawn independently for the two resolutions, since their measurement noise is independent. These augmentations introduce realistic variations in line positions and amplitudes while preserving the underlying astrophysical structure, thereby expanding the training set and promoting shift-robust, noise-tolerant learning. On the logarithmic grid of \S\ref{sec:preproc} a change of redshift is an exact translation of the spectrum, so no flux is redistributed between samples.}

    \label{fig:augmentation}
\end{figure*}

Following pre-processing, a data augmentation procedure is applied to increase the size and diversity of the training set. The pre-processed dataset consists of $2{,}858$ matched pairs. Before any augmentation, $572$ galaxies are set aside as the held-out sample, so that no realization of a held-out galaxy can enter training. For each of the remaining $2{,}286$ galaxies we generate $20$ stochastic realizations, giving $21$ rows per galaxy and $48{,}006$ training rows in total ($2{,}286$ galaxies $\times\,21$).

As illustrated in Figure~\ref{fig:augmentation}, each augmented realization is produced by applying small random redshift and flux perturbations to both the low- and high-resolution spectra. Redshift offsets are drawn from a Gaussian distribution in redshift space, $z_{\mathrm{shift}} \sim \mathcal{N}(0,\sigma_z)$ with $\sigma_z=0.3$. The offset is applied as a translation along the logarithmic grid of \S\ref{sec:preproc}, which displaces the spectrum by $\ln[(1+z_{\mathrm{new}})/(1+z)]$ in $\ln\lambda$. This introduces small continuous shifts in the apparent positions of emission features, encouraging the network to learn representations that are robust to line-position variability rather than relying on fixed grid locations.

To further emulate observational variability, Gaussian noise is added independently to the low- and high-resolution spectra. The noise amplitude is scaled to 10\% of the flux value at each wavelength. As a result, the corresponding uncertainty is increased in quadrature. Applying independent noise realizations to the input and target spectra prevents the network from simply learning to copy the input to the output, and instead forces it to recover stable high-resolution features in the presence of flux uncertainties.

The choice of 20 augmented realizations per spectrum reflects a practical balance between dataset expansion and information content. Empirically, increasing the number of augmentations beyond this point does not lead to measurable improvements in training stability or validation performance, indicating saturation in the effectiveness of additional stochastic variants. This strategy expands the training set by more than an order of magnitude while ensuring that the model continues to learn physically meaningful mappings between low- and high-resolution spectra.

\subsection{Model Architecture and Training} \label{sec:architecture}

%
%
%
%
\begin{figure*}[t]
\centering
\includegraphics[width=\textwidth]{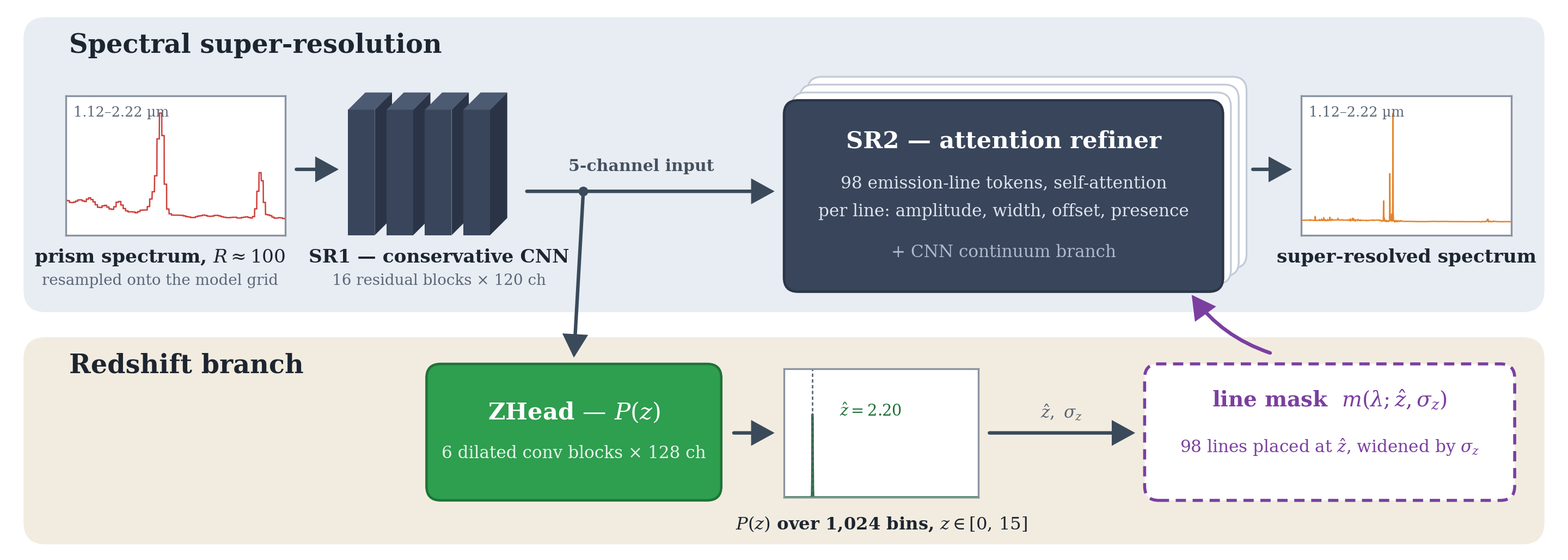}
\caption{The three-stage pipeline, with every value read from the
released checkpoints.
SR1 reconstructs a deliberately conservative super-resolved spectrum from the
prism input, together with a per-pixel uncertainty.
The redshift stage returns a probability distribution over redshift, so a
spectrum whose lines admit two identifications reports a broad distribution
instead of a confident value lying between them.
That redshift places a mask marking where lines may fall, which widens when the
redshift is uncertain.
SR2 adds a two-part correction: a line model that fits an amplitude, width and
position to each cataloged transition and may decline to place it, and a
smooth continuum term.
The insets are real output for a held-out galaxy at $z=2.198$: the prism input,
the super-resolved spectrum, and the predicted $P(z)$, with the dashed line at
the catalog redshift.}
\label{fig:sr_architecture}
\end{figure*}

Recovering high-resolution spectral structure from low-resolution observations is an intrinsically ill-posed problem. End-to-end super-resolution models trained directly from low- to high-resolution spectra tend to either oversmooth physically real features or hallucinate sharp lines that are inconsistent with any plausible spectroscopic interpretation (\citealt{ledig2017srgan}; \citealt{blau2018perception}; \citealt{cohen2018hallucinate}). To address these challenges, we adopt a staged architecture that explicitly separates coarse reconstruction, physical interpretation, and fine-scale refinement. An overview of the full three-stage pipeline is shown in Figure~\ref{fig:sr_architecture}.

The first stage (hereafter \gls{sr1}) performs a conservative super-resolution of the low-resolution spectrum onto a fixed, high-resolution wavelength grid. The goal of \gls{sr1} is not to recover the sharpest possible spectral features, but rather to reconstruct the global continuum shape and broad spectral structure in a stable and noise-robust manner. This conservatism is a property of what \gls{sr1} is trained to prioritize and should not be mistaken for the resampling of \S\ref{sec:preproc}, which is a fixed, deterministic re-binning with no free parameters that never sees a grating spectrum and cannot introduce a feature the prism does not already contain. \gls{sr1} is trained with the grating spectrum as its target, so it learns the statistical relationship between a prism spectrum and the medium-resolution spectrum of the same galaxy, and can sharpen and separate structure that resampling only carries across unchanged.

Both \gls{sr1} and subsequent stages predict wavelength-dependent uncertainties in addition to the reconstructed flux. Letting the predicted uncertainty vary freely from pixel to pixel (a heteroscedastic formulation) reflects the fact that spectroscopic noise varies across the detector band-pass and that the high-resolution training targets may themselves contain residual calibration artifacts or measurement uncertainties.

In the second stage, a dedicated redshift inference network is applied to the output of \gls{sr1}. It consumes both the \gls{sr1} reconstruction and its predicted log-variance and returns a probability distribution over a grid of 1,024 redshift bins covering the range spanned by the training sample. The redshift $\hat{z}$ passed to the next stage is the probability-weighted mean of the bins around the most probable one, and its uncertainty $\sigma_z$ is the standard deviation of the full distribution. Redshift is treated as a global latent parameter that constrains the relative spacing and identity of spectral features. \gls{sr1} is trained to convergence and frozen before the redshift network is trained at all. This arrangement demonstrates only that a usable redshift can be obtained from the coarse super-resolved spectrum; whether super-resolution \emph{improves} redshift recovery is measured in \S\ref{sec:redshift_consistency}.

Using the inferred redshift, a soft, wavelength-dependent line mask is constructed by shifting a rest-frame emission-line list into the observed frame. Each Gaussian is broadened in quadrature by $\lambda_{\rm rest}\sigma_{z}$, the wavelength displacement implied by the predicted redshift uncertainty, and capped at 0.05\,$\mu$m, so a confident redshift yields a narrow mask and an unconfident one a broad, uninformative mask. Each Gaussian is broadened in quadrature by $\lambda_{\rm rest}\sigma_{z}$, the wavelength displacement implied by the predicted redshift uncertainty, and capped at 0.05\,$\mu$m so that a very uncertain redshift does not widen the mask over the whole spectrum. This mask does not enforce the presence of specific lines; rather, it encodes physical plausibility in a probabilistic manner that can be incorporated into the refinement process.

The third stage (\gls{sr2}) refines the \gls{sr1} output by predicting a residual correction rather than a full reconstruction. Its input is a stack of five arrays on the common wavelength grid (five branches): the resampled prism spectrum, the \gls{sr1} reconstruction, the \gls{sr1} uncertainty, the line mask, and the predicted redshift repeated at every wavelength. \gls{sr2} itself is split into two sub-networks that read the same input: one models the emission lines and the other corrects the smooth continuum. This mirrors what a galaxy spectrum is, a slowly varying continuum with narrow lines on top of it, and the two corrections are added at the end.

The first branch is a physics-informed line modeler in which every cataloged line is modeled jointly with all the others (multi-head self-attention). For each of $K$ known rest-frame emission lines, the inferred redshift $\hat{z}$ is used to compute the expected observed-frame wavelength, and a local spectral window of $2W+1$ pixels is extracted from the input channels centered on that position. Each window is encoded by a small convolutional network that summarizes it on several scales at once (multi-scale adaptive pooling), and a learned vector unique to each line species (an identity embedding) is added to that summary so that lines of similar local morphology, such as the two components of a doublet, are distinguished by their physical identity.

The resulting $K$ per-line summaries (line tokens) pass through a stack of layers in which each line is compared against all the others (a Transformer encoder), so that every line can exchange information with every other, enabling the network to learn inter-line relationships such as the fixed ${\sim}1{:}3$ flux ratio of the [O\,\textsc{iii}]\,$\lambda\lambda4959,5007$ doublet, the Balmer decrement linking H$\alpha$ and H$\beta$, and correlations between ionization-sensitive line groups, which a purely local convolutional architecture cannot capture because the relevant lines may be separated by hundreds of pixels.

Each line token is then decoded into a parametric Gaussian profile. Separate linear heads predict the amplitude, log-width, and sub-pixel wavelength offset of each line, as well as a presence probability, squashed to lie between 0 and 1 (a sigmoid gate). The presence gate allows the network to suppress lines that are not supported by the data rather than forcing every known line to appear, reducing hallucination; it is trained explicitly against the lines the reference spectrum shows, so declining to place a line is a supervised decision. The full line-branch residual is reconstructed as a sparse sum of Gaussians, each deposited at the predicted position with the predicted width and gated amplitude.

The second branch is a shallow residual convolutional network that predicts a smooth continuum correction. The two branches are summed to produce the total residual prediction: $\Delta x = \Delta x_{\mathrm{line}} + \Delta x_{\mathrm{CNN}}$, and the final super-resolved spectrum is $x_{\mathrm{SR2}} = x_{\mathrm{SR1}} + \Delta x$. A separate convolutional head predicts the wavelength-dependent uncertainty, carried as a log-variance, from the input channels.

Refinement is guided by loss terms that emphasize high-frequency spectral components corresponding to narrow lines and sharp features. High-pass filtered components of the refined spectrum are encouraged to match those of the high-resolution target preferentially within the line mask, while regions outside the mask are biased toward smoothness and minimal deviation from the coarse reconstruction. A further term compares the integrated flux of each reconstructed line with the same quantity measured on the reference, since matching the shape of a line does not by itself guarantee matching the flux it carries.

Line-shape residuals are weighted uniformly within the line mask, \emph{not} by inverse variance; the continuum term retains inverse-variance weighting. Photon noise scales with flux, so the brightest pixels of an emission line are also its noisiest, and inverse-variance weighting would concentrate the line term on the faint wings: the faintest tenth of in-line pixels would receive $34\times$ the weight of the brightest tenth, leaving the term nearly blind to the cores that carry the flux.

All three components of the pipeline are implemented as one-dimensional neural networks operating along the wavelength axis. \gls{sr1} and the \gls{cnn} branch of \gls{sr2} use convolutional architectures that reflect the inherently ordered and local nature of spectroscopic data, while the line-attention branch of \gls{sr2} uses a Transformer encoder that captures long-range inter-line correlations. The full three-stage pipeline totals 3,101,705 trainable parameters (\gls{sr1} 1,391,042; the redshift head 710,145; \gls{sr2} 1,000,518).

The initial super-resolution network (\gls{sr1}) adopts a residual convolutional architecture composed of a shallow input projection followed by a stack of residual blocks with small convolutional kernels. The final layers predict both the reconstructed flux and a wavelength-dependent uncertainty, carried as a log-variance, enabling uncertainty estimation.

Training proceeds sequentially: \gls{sr1} is trained first and frozen, followed by the redshift inference network, and finally the \gls{sr2} refinement stage. \gls{sr1} and the redshift head are held fixed throughout \gls{sr2} training. \gls{sr2} is trained on the redshifts the head predicts, rather than catalog values.

\begin{figure*}[ht!]
    \centering
    \includegraphics[width=\linewidth]{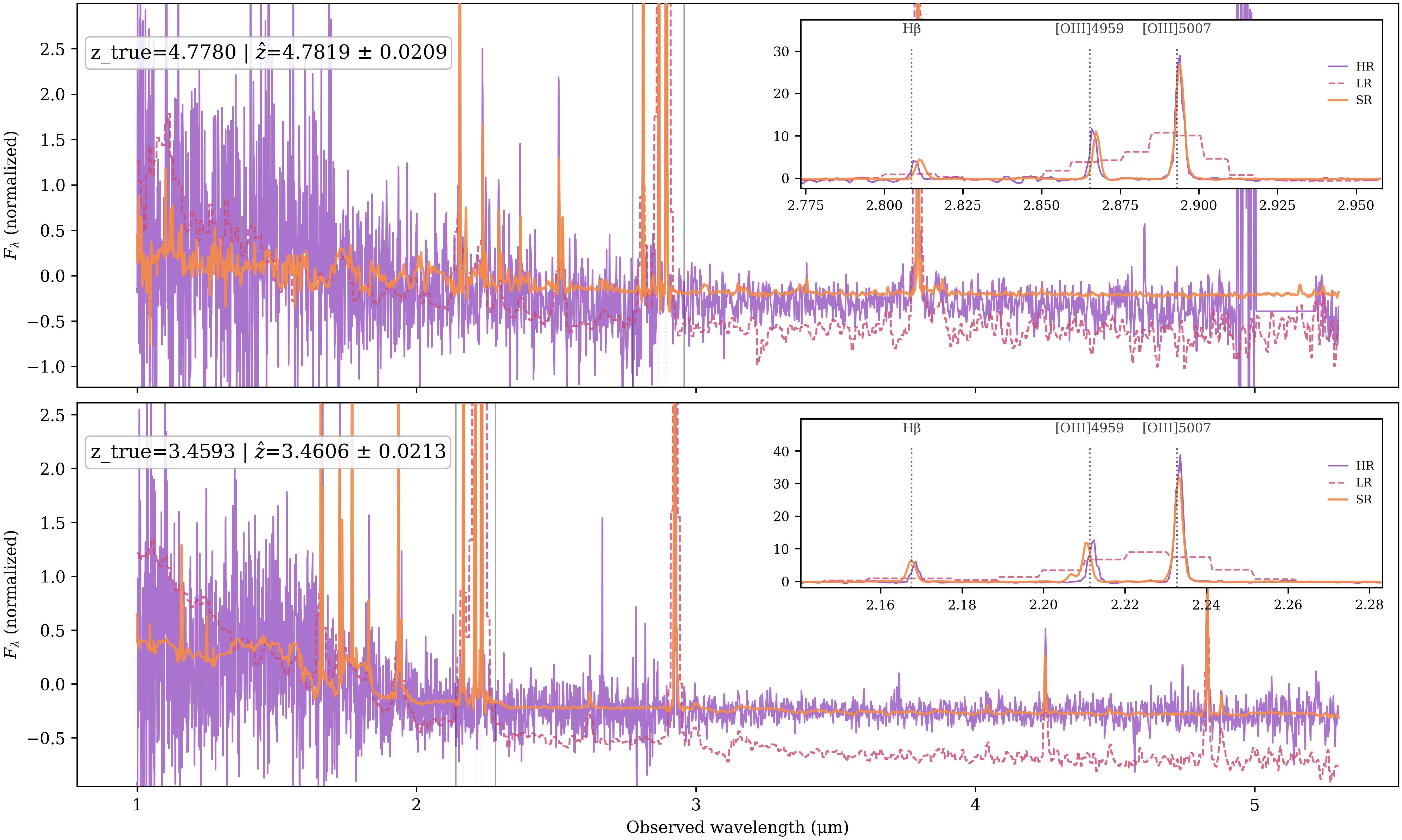}
    \caption{
    Two example galaxies from the 20\% held-out evaluation set, neither
    used during training. Each panel shows the low-resolution \textit{prism}
    input (red dashed line), the high-resolution target (purple solid line), and
    the super-resolved output (orange solid line) over the full
    $1.0$--$5.3$\,$\mathrm{\mu m}$ grid. The boxed label gives the catalog
    redshift and the value returned by the redshift stage. The inset zooms on
    H$\beta$ and the [O\,\textsc{iii}]\,$\lambda\lambda4959,5007$ doublet,
    which are unresolved at prism resolution, but separated in \gls{hr} and \gls{sr2} predictions.}
    
    \label{fig:generated_spectra}
\end{figure*}
\section{Results}\label{sec:results}
We evaluate the performance of our super-resolution model on a 20\% held-out subset of the \gls{jades} sample that was never seen during training, comprising the real observed spectra of $572$ galaxies. Below, we present representative reconstructions (\S\ref{sec:generated_spectra}), assess the wavelength-dependent fidelity of the model through residual analysis (\S\ref{sec:residual_maps}), quantify the improvement in emission-line recoverability and ask whether it yields usable line fluxes and ratios (\S\ref{sec:line_recovery}), and evaluate redshift consistency as a downstream diagnostic of recovered information content (\S\ref{sec:redshift_consistency}).

\subsection{Generated high-resolution spectra}\label{sec:generated_spectra}

Figure~\ref{fig:generated_spectra} presents two examples from the 20\% held-out evaluation set, selected to illustrate the model's ability to deblend emission lines that are unresolved at prism resolution. In both cases, the low-resolution prism spectrum (red dashed line) shows a single broad feature in the [O\,\textsc{iii}] region, with a very low S/N H$\beta$, and the [O\,\textsc{iii}]\,$\lambda4959$, and [O\,\textsc{iii}]\,$\lambda5007$ are fully blended. The super-resolved output (orange solid line) successfully separates these lines into distinct peaks at the correct observed-frame wavelengths, closely matching the medium-resolution reference spectrum (purple solid line), as shown in the inset panels. Some residual fluxes especially at the line peaks are visible. They are the under-estimation of emission-line flux discussed in detail in \S\ref{sec:line_recovery}. Overall, it should be noted that visually separated doublet does not necessarily yet yield a precise $\lambda5007/\lambda4959$ ratio.

\subsection{Residual Maps}\label{sec:residual_maps}

\begin{figure*}[ht!]
    \centering
    \includegraphics[width=\linewidth]{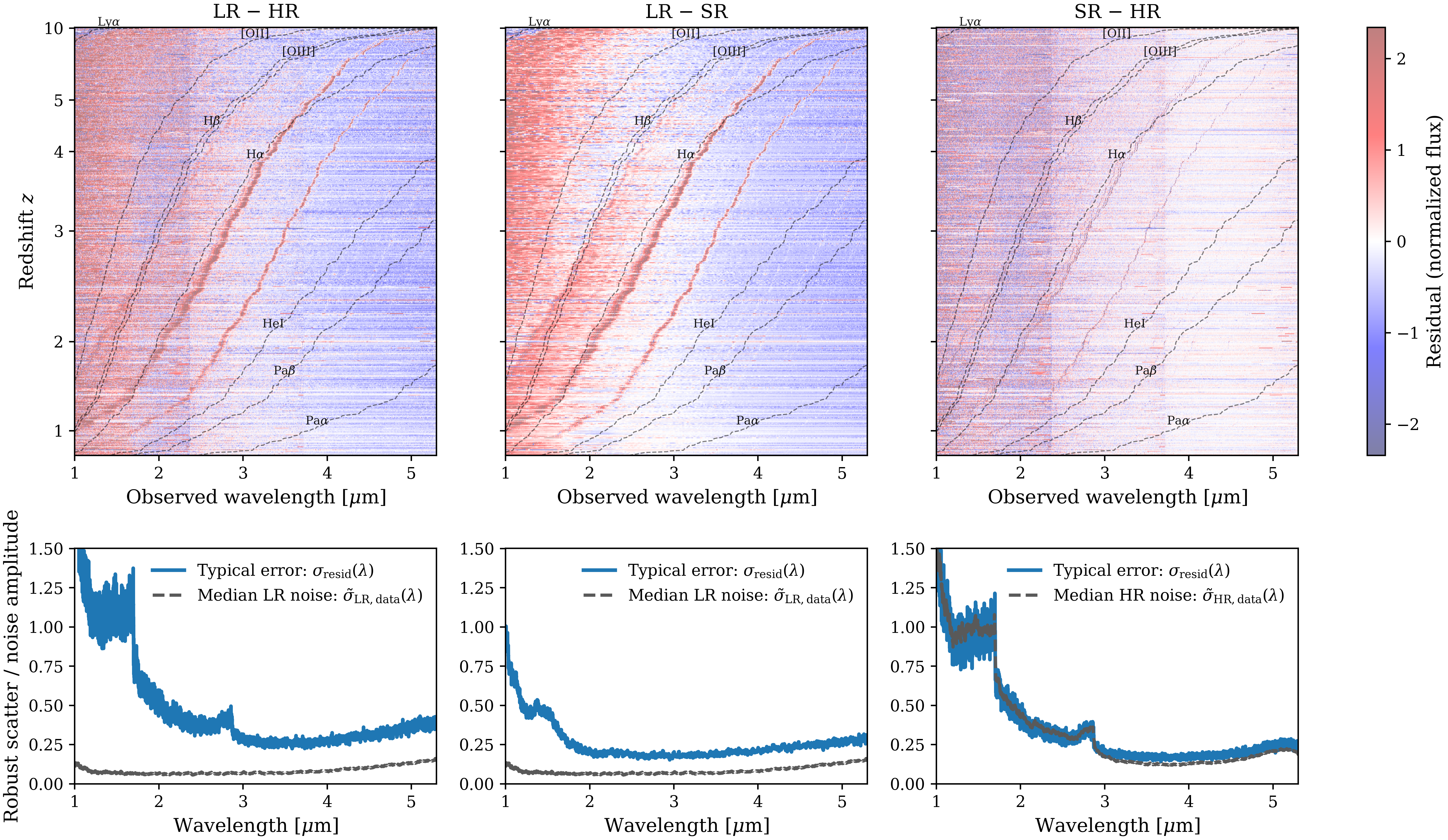}
    \caption{
    Residual structure and comparison to data noise.
    \textit{Top row:}
    Residual maps for the same set of galaxies, sorted by true redshift, showing (from left to right) the difference between the resampled low-resolution spectra and high-resolution spectra (\gls{lr}--\gls{hr}), low-resolution spectra and the \gls{sr2} reconstruction (\gls{lr}--\gls{sr2}), and the \gls{sr2} reconstruction and high-resolution spectra (\gls{sr2}--\gls{hr}). Residuals are shown in normalized flux units and share a common color scale. Coherent wavelength-dependent structure in the \gls{lr}--\gls{hr} residuals indicates that low-resolution spectra retain correlated information about high-resolution features, while the \gls{sr2}--\gls{hr} residuals show only faint line tracks, far weaker than in the \gls{lr}--\gls{hr} map. The bottom row quantifies the residual against the noise.
    \textit{Bottom row:}
    For each residual map, the solid blue curve shows the typical reconstruction error as a function of wavelength, quantified as the robust scatter (MAD-based standard deviation) of the residuals across galaxies, $\sigma_{\rm resid}(\lambda)$. The dashed gray curve shows the median wavelength-dependent noise amplitude of the input data, $\tilde{\sigma}_{\rm data}(\lambda)$, estimated independently from the spectra via high-frequency residuals after continuum smoothing. For the \gls{lr}--\gls{hr} and \gls{lr}--\gls{sr2} panels, the noise estimate is derived from the low-resolution data, while for the \gls{sr2}--\gls{hr} panel it is derived from the high-resolution data. Agreement between $\sigma_{\rm resid}(\lambda)$ and $\tilde{\sigma}_{\rm data}(\lambda)$ indicates noise-limited performance, whereas smaller residual scatter implies implicit denoising by the model.
    }
    \label{fig:residual_maps}
\end{figure*}

\begin{figure*}[ht!]
    \centering
    \includegraphics[width=\linewidth]{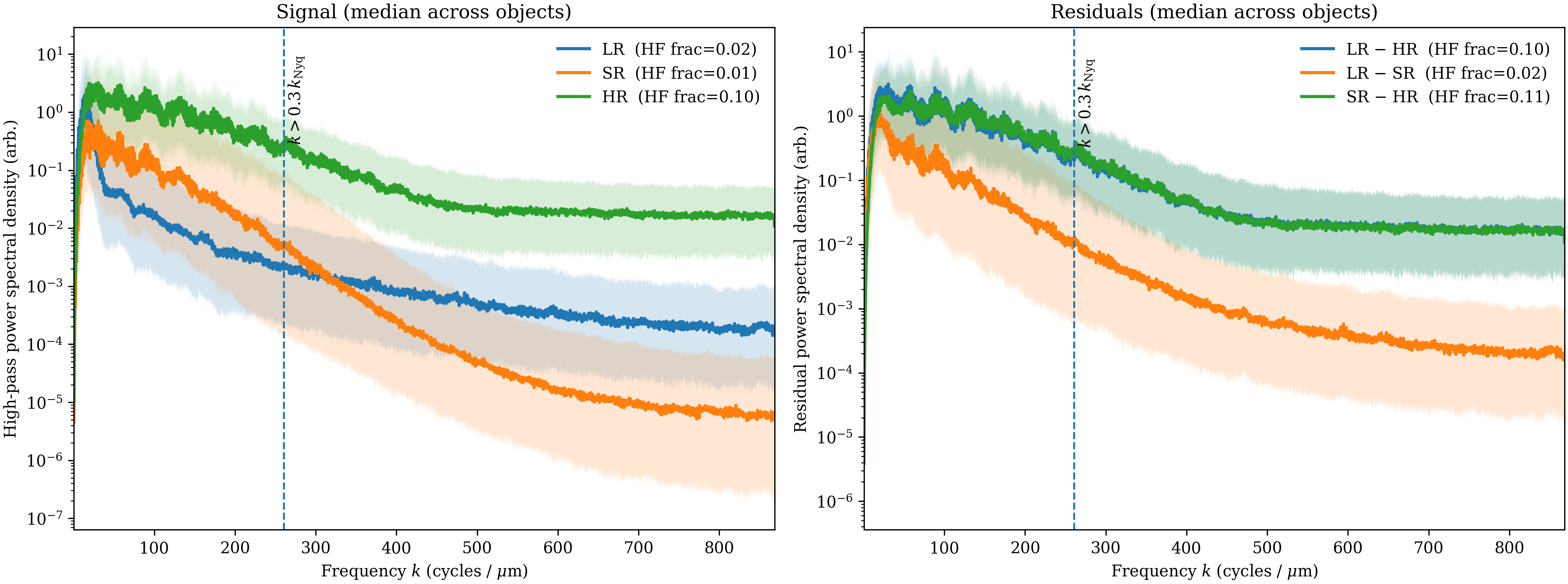}
    \caption{
    Frequency--domain diagnostics of the super--resolution performance.
    \emph{Left:} \Gls{psd} of the high--pass component of the spectra, computed by subtracting a smoothed continuum (window size = 101 pixels) and taking the median across objects.
    Solid lines show the median \gls{psd}, while shaded regions indicate the 16th--84th percentile range across the sample.
    The \gls{hr} spectra carry by far the most small--scale structure above $0.3\,k_{\mathrm{Nyq}}$, while the \gls{sr} spectra add structure at intermediate scales; the text quantifies both.
    \emph{Right:} \gls{psd} of residuals between different spectral products, shown as median curves with 16th--84th percentile ranges.
    Both members of each pair are carried on the high--resolution sampling by flux--conserving resampling, so neither is degraded by the gridding.
    The vertical dashed line marks $0.3\,k_{\mathrm{Nyq}}$, separating intermediate--scale structure from noise--dominated frequencies.
    }
    \label{fig:PSD}
\end{figure*}

To assess the fidelity of the reconstructed spectra, we examine the residuals between different spectral products and the high-resolution reference spectra. Because galaxies span orders of magnitude in continuum level and emission-line strength, all residual statistics are computed in normalized flux units, using the same per-object normalization applied during training. This removes astrophysical dynamic range from the analysis and places all spectra on a common scale, ensuring that the residuals reflect model performance rather than intrinsic flux diversity.

Figure~\ref{fig:residual_maps} presents this analysis across the full 20\% held-out evaluation set. The top row shows two-dimensional residual maps for three comparisons: the low-resolution input minus the high-resolution reference (\gls{lr}$-$\gls{hr}), the low-resolution input minus the super-resolved output (\gls{lr}$-$\gls{sr2}), and the super-resolved output minus the high-resolution reference (\gls{sr2}$-$\gls{hr}). The \gls{lr}$-$\gls{hr} map reveals strong, coherent residual structure tracing the redshift-dependent positions of emission lines. The \gls{lr}$-$\gls{sr2} map shows similar structure, indicating that the model successfully introduces emission-line features that were missing in the \gls{lr} input. The \gls{sr2}$-$\gls{hr} map is close to featureless by comparison.

The bottom row compares the wavelength-dependent reconstruction error, quantified as the robust scatter of the residuals across galaxies ($\sigma_\mathrm{resid}(\lambda)$), with the median noise amplitude of the \gls{hr} data ($\tilde{\sigma}_\mathrm{HR,data}(\lambda)$). For the \gls{sr2}$-$\gls{hr} comparison the ratio $\sigma_\mathrm{resid}/\tilde{\sigma}_\mathrm{HR,data}$ is $0.84$, $0.94$ and $0.92$ in the $1.0$--$1.5$, $1.5$--$2.0$ and $2.0$--$3.0$\,$\mu$m bands and rises to $1.25$ and $1.15$ at $3$--$4$ and $4$--$5.3$\,$\mu$m, so the reconstruction is close to noise-limited over most of the range and exceeds the reference noise beyond $3$\,$\mu$m. A ratio below one at the blue end does not mean the reconstruction is accurate there. Where the reference is as noisy as it is below $1.5$\,$\mu$m, any smooth prediction lies closer to it than its own noise, whether or not the features it draws are right; the line-flux comparison of \S\ref{sec:line_recovery} is the test that tells the two apart.

The band ratios above include all galaxies at each wavelength, and because each line falls at a different observed wavelength for each redshift, the pixels on emission lines are a small minority at any one wavelength. Measured line by line instead, within $\pm500$\,km\,s$^{-1}$ of each cataloged line detected in the reference and compared with a continuum window $2000$--$4000$\,km\,s$^{-1}$ away in the same spectrum, the median $|$\gls{sr2}$-$\gls{hr}$|$ is larger on the line than beside it by factors of $6.3$ for [O\,\textsc{iii}]\,$\lambda5007$, $6.2$ for H$\alpha$, $3.4$ for [O\,\textsc{ii}]\,$\lambda3727$, $3.1$ for H$\beta$, $2.3$ for H$\gamma$ and $1.9$ for H$\delta$, and \gls{sr2} lies below the reference on every line. The faint diagonal ridges that remain in the \gls{sr2}$-$\gls{hr} map are therefore real: they are the under-estimation of line flux quantified in \S\ref{sec:line_recovery}, far weaker than the ridges of the \gls{lr}$-$\gls{hr} map.

We complement the spatial-domain residual analysis with a frequency-domain diagnostic. Figure~\ref{fig:PSD} shows the \gls{psd} of the high-pass component of the spectra, obtained by subtracting a smoothed continuum, for the \gls{lr}, \gls{sr}, and \gls{hr} products. The dashed line at $0.3\,k_{\mathrm{Nyq}}$ separates intermediate-scale structure from the frequencies where noise dominates. Above it the \gls{hr} spectra carry by far the most fine-scale structure: their median power is $54$ times that of the \gls{lr} spectra and $823$ times that of the \gls{sr} spectra.

The right panel shows the \gls{psd} of the residuals between products. Above $0.3\,k_{\mathrm{Nyq}}$ the \gls{lr}$-$\gls{sr} residuals lie a factor of $46$ below the \gls{lr}$-$\gls{hr} residuals, so at these frequencies the reconstruction stays close to the spectrum it was given, while the \gls{sr}$-$\gls{hr} residual power is $0.97$ times the \gls{lr}$-$\gls{hr} value, meaning the finest scales in the grating spectra are not recovered. This is the limit the line-flux measurement of \S\ref{sec:line_recovery} quantifies, seen in the frequency domain.

\subsection{Emission-Line Recovery}\label{sec:line_recovery}

\begin{figure*}[ht!]
    \centering
    \includegraphics[width=\linewidth]{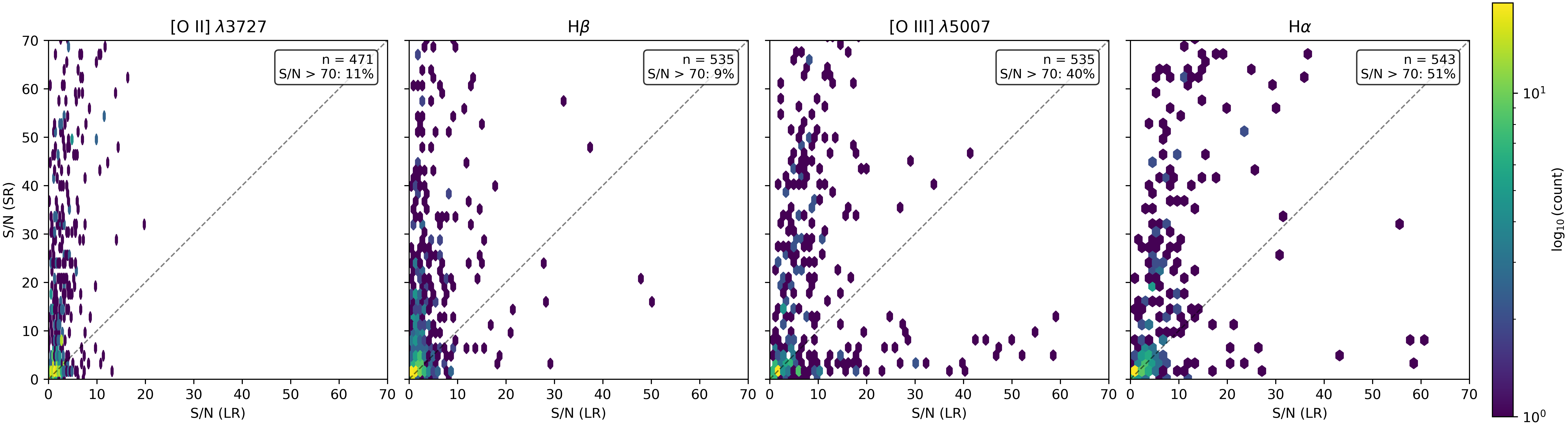}
    \caption{Comparison of emission-line S/N between \gls{lr} and \gls{sr} spectra for [O\,\textsc{ii}]\,$\lambda3727$, H$\beta$, [O\,\textsc{iii}]\,$\lambda5007$, and H$\alpha$. Each point represents a single galaxy from the 20\% held-out evaluation set. Each panel's box reports $n$, the number of galaxies shown in that panel (those in whose spectra the line is detected). The fraction measured above $\mathrm{S/N} = 70$ is also reported. S/N values are obtained from Gaussian line fits (Equation~\ref{eq:line_fit}), with the local continuum estimated from sidebands adjacent to each line. The dashed line marks the one-to-one relation. Points above this line indicate improved S/N in the super-resolved spectrum. The color scale indicates the density of points in $\log_{10}$ counts.}
    \label{fig:sn_modification}
\end{figure*}

To quantify the recoverability of individual emission features, we fit a Gaussian line model to the low-resolution \textit{prism} spectra, the super-resolved outputs and the medium-resolution reference. For each emission line in a galaxy, we compute the expected observed wavelength from the catalog redshift and extract a window of $\pm0.25$\,$\mathrm{\mu m}$ around that position, with the local continuum measured from sidebands beyond a $\pm0.05$\,$\mathrm{\mu m}$ core. Within this window, we model the observed flux as
\begin{align}
f(\lambda) = C_{0} + C_{1}\lambda + A \exp\!\left[-\frac{(\lambda - \mu)^2}{2\sigma^2}\right],
\label{eq:line_fit}
\end{align}
where $A$ is the line amplitude, $\mu$ the centroid, $\sigma$ the width, and $(C_{0},C_{1})$ describe the local continuum level and slope (\citealt{haghjoo_thesis_mcgill}). The emission-line S/N is $A/\sigma_\mathrm{cont}$, where $\sigma_\mathrm{cont}$ is the robust scatter of the flux in the sidebands, so each line is measured against the noise of the neighboring continuum.

Using this procedure, we compare the S/N measured from the low-resolution \textit{prism} spectra with that measured from the super-resolved outputs across the full evaluation set. Figure~\ref{fig:sn_modification} summarizes this comparison for four key diagnostic lines: [O\,\textsc{II}]\,$\lambda3727$, H$\beta$, [O\,\textsc{III}]\,$\lambda5007$, and H$\alpha$. The median per-galaxy gain is a factor of $5.68$ for [O\,\textsc{II}]\,$\lambda3727$, $3.31$ for H$\beta$, $5.26$ for [O\,\textsc{III}]\,$\lambda5007$ and $5.32$ for H$\alpha$, with $84.3\%$, $77.0\%$, $79.3\%$ and $85.5\%$ of galaxies respectively improving. The size of the gain is not set by how faint the line is in the prism: [O\,\textsc{II}] and H$\beta$ are the two faintest lines there, with median prism S/N of $2.5$ and $2.3$, yet [O\,\textsc{II}] shows the largest gain of the four and H$\beta$ the smallest.

The remaining $15$--$23\%$ of galaxies fall near or below the one-to-one line for two reasons. \textit{1. Faintness}: their median reference S/N is lower than that of the galaxies above the line, by a factor of $0.39$ for H$\alpha$ up to $0.60$ for [O\,\textsc{II}]. \textit{2. Accuracy of the Inferred Redshift}: for [O\,\textsc{III}]\,$\lambda5007$, the median velocity offset between the inferred and the catalog redshift is $2.4$ fitted line widths for the galaxies above the line and $7.7$ for those below it (the median fitted width of the reconstructed line being $172\,\mathrm{km\,s^{-1}}$), and the other three lines behave the same way.

The gain has two sources, a stronger fitted line and a smoother continuum beside it, and they are not equal. For H$\alpha$ over the held-out set the fitted amplitude rises by a factor of $1.32$ while the continuum scatter beside the line falls by a factor of $4.53$, so most of the apparent improvement comes from the reconstruction being smooth, not from the line being stronger.

\begin{figure*}[ht!]
    \centering
    \includegraphics[width=\linewidth]{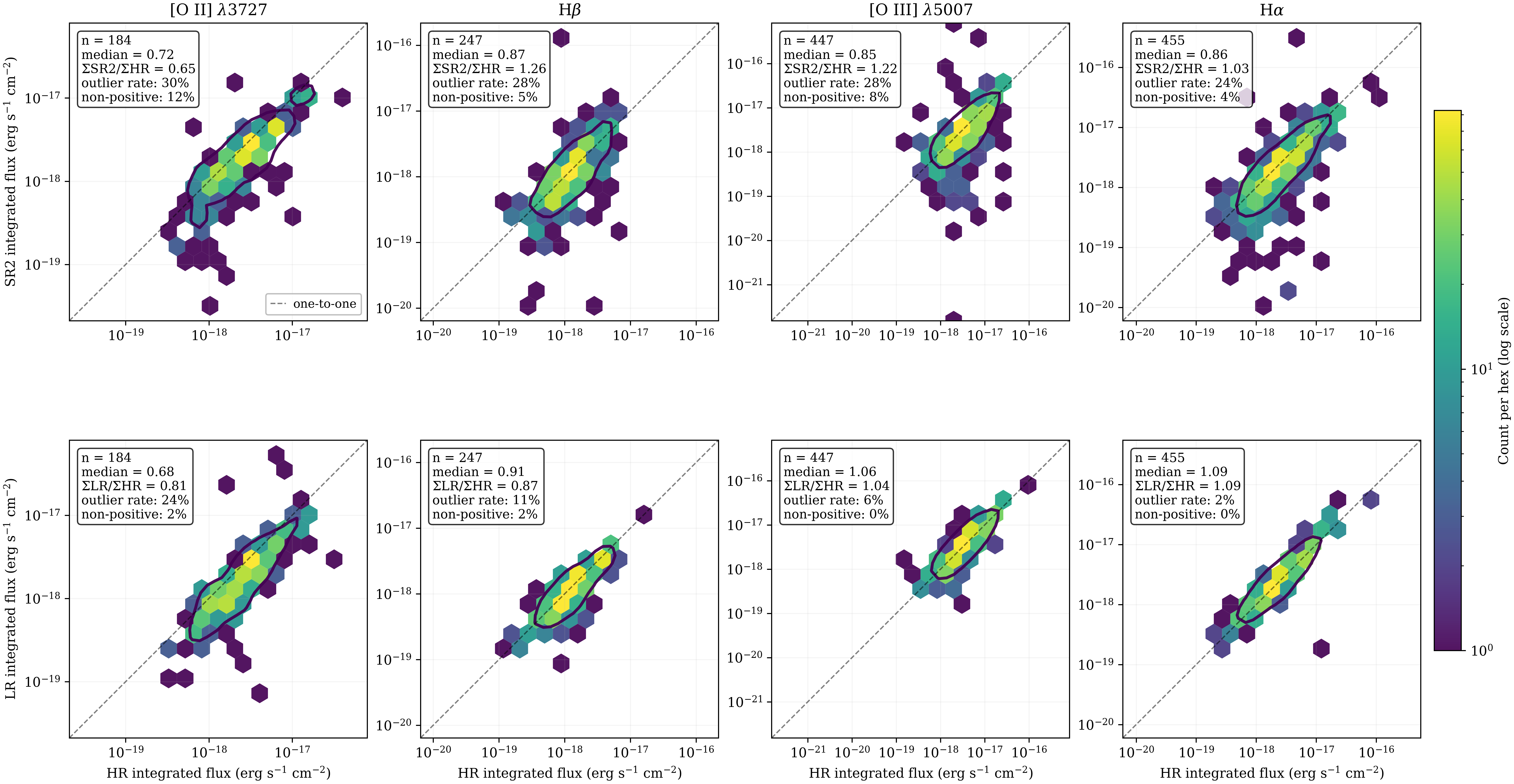}
    \caption{Integrated emission-line flux against the medium-resolution reference for the same four lines as Figure~\ref{fig:sn_modification}. Each flux is the integral of a fitted Gaussian: Equation~\ref{eq:line_fit} for [O\,\textsc{II}] and H$\alpha$, and a joint fit of H$\beta$, $\lambda4959$ and $\lambda5007$ with a shared width for the other two. The upper row is the \gls{sr2} reconstruction and the lower row the prism it was given as input, both drawn on the same axes. The contour encloses the $68\%$ highest-density region. Each box reports $n$ (number of lines detected at $\mathrm{S/N} \geq 5$ in the reference), the median per-line ratio to the reference, the ratio of summed fluxes, the fraction of lines wrong by more than a factor of two either way, and the fraction that cannot appear on a logarithmic axis (fits of zero or negative amplitude, counted as outliers).}

    \label{fig:line_flux}
\end{figure*}

We can also investigate the integrated line flux measured against the reference (Figure~\ref{fig:line_flux}). The flux of each line is the integral of a fitted Gaussian, $F = \sqrt{2\pi}\,A\,\sigma$, so every spectrum is measured through its own line profile. [O\,\textsc{II}]\,$\lambda3727$ and H$\alpha$ are fit with Equation~\ref{eq:line_fit}. H$\beta$, [O\,\textsc{III}]\,$\lambda4959$ and [O\,\textsc{III}]\,$\lambda5007$ lie within $9000$\,km\,s$^{-1}$ of one another, and the prism, whose line width at these wavelengths has a median \gls{fwhm} of $4800$\,km\,s$^{-1}$, does not separate $\lambda4959$ from $\lambda5007$ in $84\%$ of galaxies, or H$\beta$ from the doublet in $45\%$. The three are therefore fit jointly as Gaussians of a shared width, the model of the doublet test below, so that the flux of $\lambda4959$ is not assigned to its neighbors. A line enters the comparison when the reference detects it in emission at $\mathrm{S/N} \geq 5$. The median \gls{sr2}-to-reference flux ratio is $0.72$ for [O\,\textsc{II}]\,$\lambda3727$, $0.87$ for H$\beta$, $0.85$ for [O\,\textsc{III}]\,$\lambda5007$ and $0.86$ for H$\alpha$ ($0.83$ over the four lines together). The per-line scatter demonstrates that $24$--$30\%$ of individual lines are outliers by a factor of two (Figure~\ref{fig:line_flux}, upper panel). The same measurement on the prism input gives median ratios of $0.68$, $0.91$, $1.06$ and $1.09$ for the four lines, with outlier rates of $24\%$, $11\%$, $6\%$ and $2\%$ (Figure~\ref{fig:line_flux}, lower panel; the excess for H$\alpha$ is consistent with the [N\,\textsc{II}] lines that a single Gaussian at prism resolution takes in). Therefore, the \gls{sr2} line fluxes are under-estimated by $13$--$28\%$ in the median at the present training-set size.

The deficit depends on the inferred redshift, at which the refinement stage places its line model; Lines whose redshift is recovered to better than $500$\,km\,s$^{-1}$ ($57.9\%$ of the lines in Figure~\ref{fig:line_flux}) reach a median flux ratio of $0.90$ with $18\%$ outliers, while the remainder reach $0.70$ with $39\%$.

The [O\,\textsc{III}]\,$\lambda\lambda4959,5007$ ratio is fixed by atomic physics at $2.98$. Fitting three Gaussians of a shared width at the positions of H$\beta$, $\lambda4959$ and $\lambda5007$ in the galaxies whose $\lambda5007$ is detected at $\mathrm{S/N} \geq 20$ in the reference, the ratio is recovered to within $20\%$ for $85.8\%$ of galaxies from the grating, $39.3\%$ from the prism and $32.4\%$ from \gls{sr2}. The fitted widths, $168$, $1848$ and $365$\,km\,s$^{-1}$, give $R \simeq c/(2.3548\sigma) \simeq 760$, $70$ and $350$, lower limits since each includes the intrinsic width of the line.

\subsection{Redshift Consistency as a Downstream Diagnostic}\label{sec:redshift_consistency}

Although the primary goal of this work is not redshift estimation, we use redshift consistency as an informative downstream diagnostic to assess how much physically relevant information is recovered by the \gls{sr} spectra. To this end we train the same redshift network on four representations of the same galaxies: the native \gls{lr} prism spectra, the \gls{sr1} output, the \gls{sr2} output, and the \gls{hr} grating reference, with the \gls{jades} catalog redshifts taken as ground truth (\S\ref{sec:data}). All four arms come from a single training routine invoked with one argument changed, so architecture, loss, optimizer, schedule, data split and random seed are identical; the only difference is that the \gls{sr} arms also receive the uncertainty predicted by SR1. Every arm is trained and evaluated on the same $572$ held-out galaxies.

A gain in spectral resolution is not free at fixed exposure time: a spectrograph spreads a fixed photon budget over more pixels as its resolving power rises, so sensitivity per resolution element trades against resolution. Measuring the integrated S/N of each cataloged transition in a velocity window matched to each instrument's resolution, the prism detects a median of $21$ lines per galaxy at $\mathrm{S/N} \geq 3$ against the gratings' $16$. The relevant question is not whether super-resolution beats the prism outright, but whether it can approach the grating's precision ($|\Delta z|/(1+z) \approx 0.0008$) without inheriting its detection losses.

\begin{figure*}[ht!]
    \centering
    \includegraphics[width=\linewidth]{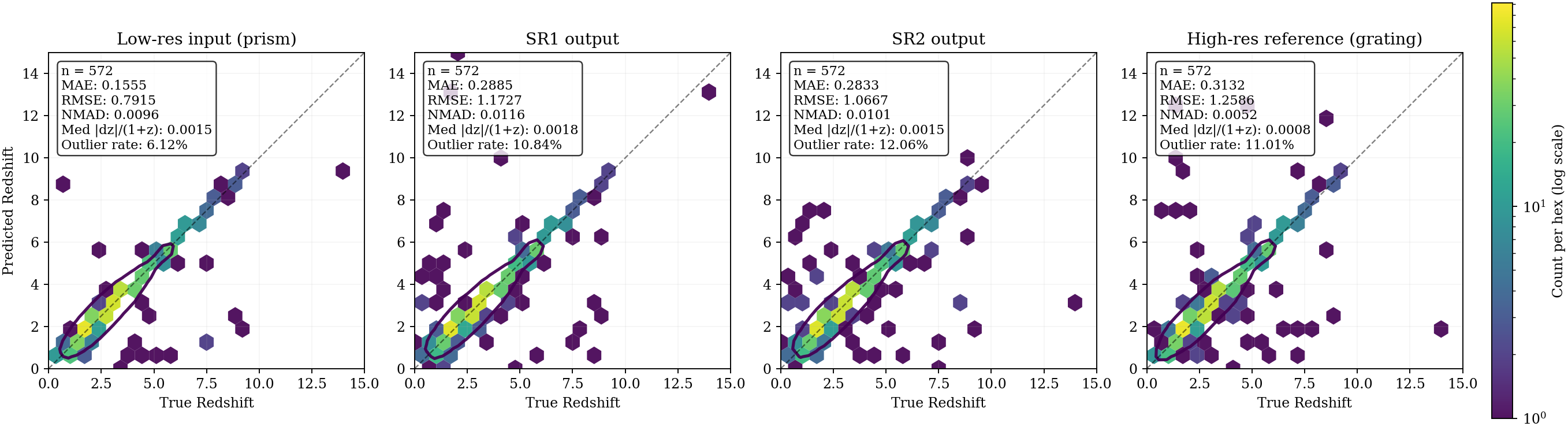}
    \caption{
    Comparison of predicted and true redshifts from identical redshift networks trained on four representations of the same galaxies: the low-resolution prism input, the \gls{sr1} output, the \gls{sr2} output, and the \gls{hr} grating reference. All panels share axis limits and color normalization and show the one-to-one relation (dashed line). The hexagonal density shows the $572$ held-out galaxies and the contour encloses the $68\%$ highest-density region. Each panel's box reports the mean and root-mean-square absolute error on $\Delta z = z_\mathrm{pred} - z_\mathrm{true}$, the normalized median absolute deviation $1.4826\,\mathrm{med}|\Delta z|$ taken on the raw residual, the median of $|\Delta z|/(1+z)$, and the outlier rate (the fraction with $|\Delta z|/(1+z) > 0.15$).
    }
    \label{fig:redshift_comparison_2}
\end{figure*}

Figure~\ref{fig:redshift_comparison_2} compares predicted against true redshift for the four inputs. A redshift estimate has to do two things: identify which transition a feature is, and measure where that feature sits. Sharper lines help the second, and the grating locates an identified line $1.8$ times more precisely than the prism. Brighter lines help the first, and the prism, which collects the same photons into fewer pixels, identifies the line system correctly more often, with catastrophic outliers for $6.1\%$ of galaxies against the grating's $11.0\%$. The prism and grating spectra are both real observations of the same galaxies, so this balance is a property of the two data sets and not of any model. Super-resolution does not change it; \gls{sr2} matches the prism's precision, and \gls{sr1} falls behind its own input.

The \gls{sr2} arm is not an independent measurement, and its numbers must be read accordingly. Its line branch draws Gaussians at the redshift the pipeline has already inferred, so measuring a redshift from an \gls{sr2} spectrum largely reads that estimate back rather than testing it; the two agree to better than $1\%$ for $86\%$ of galaxies. \gls{sr2} therefore inherits the redshift stage's mistakes wholesale: it is a catastrophic outlier for $3.1\%$ of the galaxies whose line system that stage identified correctly and for $85.5\%$ of those it did not. What the outlier rate measures in this arm is the redshift stage, not the refinement stage.

We emphasize that these results should not be interpreted as optimized redshift measurements, nor as representative of a complete galaxy population.

\section{Discussion}\label{sec:discussions}

In this work, we have presented a proof of concept for deep-learning-based spectral super-resolution of galaxy spectra, demonstrating that a staged neural network trained on paired \gls{jwst}/\gls{nirspec} observations can recover emission-line features that are entirely blended at prism resolution. The model achieves close to noise-limited residuals over most of the spectral range, systematically improves the S/N of key diagnostic lines, and separates features such as the [O\,\textsc{III}] doublet that are entirely blended in the prism input. However, at this state of the model and with the current training dataset, the network does not yet yield more accurate doublet flux ratios than direct profile fitting of the prism achieves. While these results are encouraging, the current framework has important limitations that define the path for future development.

We adopt the \gls{nirspec} prism ($R\sim100$) and medium-resolution grating modes ($R\sim1000$) as our low- and high-resolution pairs, rather than targeting the highest \gls{nirspec} resolving power ($R\sim2700$). This choice is driven by data availability: truly high-resolution \gls{nirspec} observations exist, but only over limited sky coverage, such as the \gls{nirspec} Wide \gls{gto} Survey (\citealt{wide_gto}). In contrast, \gls{jades} provides a uniquely deep and homogeneous medium-resolution dataset over a significantly larger area, enabling statistically robust training. Our results therefore reflect the practical ceiling imposed by the available reference set. With future wide-area surveys obtained at $R\sim2700$, the same framework could be retrained to deliver substantially sharper reconstructions. More broadly, the quality, size, and diversity of the training data are the primary bottleneck for this approach: improvements in the model will follow directly from improvements in the available paired spectroscopic samples.

We train on real matched pairs instead of mock spectra forward-modeled through the two instrument configurations. A mock can only contain the physics its templates encode, so a network trained on one inherits the library's assumptions about line ratios, kinematics and stellar populations, whereas real pairs carry the instrument's correlated noise and calibration residuals in both resolutions. The cost is that our reference is itself an observation of finite signal-to-noise, which we account for by predicting a wavelength-dependent uncertainty at every stage and by reporting all metrics against the observed medium-resolution spectra.

Because no emission-line significance cut is applied (\S\ref{sec:preproc}), every number in \S\ref{sec:results} is measured on a sample that includes weak-line and continuum-dominated galaxies. However, the targeting of \gls{jades} itself favors actively star-forming systems (\S\ref{sec:data}).

Several further limitations of the current framework should be noted. The training sample of $2{,}286$ galaxies is small by machine-learning standards, and augmentation multiplies rows, not independent objects. The model's ability to generalize to galaxy populations that are underrepresented in the \gls{jades} sample, such as extremely dusty systems, low-mass galaxies, or sources at the highest redshifts, has not been established. The model predicts wavelength-dependent uncertainties, but we have not yet performed a detailed calibration analysis to assess whether these uncertainties are statistically well-calibrated. Performance degrades at the bluest wavelengths ($\lesssim 1.5\,\mu$m), where both the prism input and the \gls{hr} reference spectra have low S/N; this is a data limitation rather than a model failure, but it restricts the applicability of the current framework to rest-frame \gls{uv} diagnostics at lower redshifts. Finally, we note that this approach should be understood as applying a learned prior from the training data rather than recovering information that is truly present in the low-resolution input. The model cannot reconstruct features that have no statistical relationship with the prism data; its power lies in exploiting correlations learned from the paired sample. How this framework compares with classical deconvolution, in fidelity and in computational cost, is deferred to a companion paper (Haghjoo et al., in preparation) that benchmarks it against seven classical algorithms on the same \gls{jades} sample.

Despite these limitations, the framework has significant potential for both galaxy evolution and cosmological studies. High-resolution spectra are essential for refining fundamental scaling relations such as the star-formation rate--stellar mass relation, the mass--metallicity relation, and the \gls{bpt} diagnostic diagram. Current \gls{jwst} studies demonstrate that limited sample sizes and resolution effects bias these relations toward bright, spectrally resolved galaxies (e.g., \citealt{clarke2024star, laseter2024jades, nakajima2023jwst, scholte2025jwst}). By enabling high-resolution diagnostics for statistically larger samples, our approach has the potential to reduce these biases. For example, blended line measurements can blur the separation between star-forming and \gls{agn} populations on the \gls{bpt} plane (\citealt{agn_misclass}), and empirical corrections for the [N\,\textsc{ii}]/H$\alpha$ ratio introduce substantial scatter and systematic uncertainty (\citealt{faisst2018empirical}). A model that directly recovers these blended features from prism data could eliminate the need for such approximations, enabling cleaner classification and more accurate physical parameter estimates across large samples.

Improved emission-line S/N from super-resolved spectra also has direct consequences for cosmological analyses. More precise and unbiased redshift measurements strengthen galaxy clustering studies, while an expanded sample of galaxies with well-measured H$\alpha$ increases the statistical power of large-scale structure analyses, tightening constraints on cosmological parameters and the growth of structure (e.g., \citealt{cosmology_1}). By learning physically consistent line profiles from paired \gls{jwst} data, the model could in principle be applied to large-area grism surveys without requiring additional follow-up spectroscopy, effectively producing high-resolution spectral information at a fraction of the observational cost.

As discussed in \S\ref{intro} and \S\ref{sec:data}, the broad wavelength coverage of the \gls{jades} training data makes this framework readily extensible to low-resolution grism spectroscopy from \textit{Euclid} and the \textit{Roman Space Telescope}. However, applying the model to these datasets will require careful attention to domain shift: differences in instrumental noise properties, spectral resolution profiles, and wavelength coverage between \gls{nirspec} and grism spectrometers may degrade out-of-the-box performance. Addressing this will likely require some combination of fine-tuning on instrument-specific simulations or small calibration samples, and potentially domain adaptation techniques.

Looking further ahead, several avenues could substantially improve the framework. Incorporating broadband photometry as an additional input channel would provide complementary constraints on continuum shape, stellar mass, and redshift that are not available from the low-resolution spectrum alone. The recent development of foundation models for astronomical data offers a promising direction here, as pretrained representations of galaxy spectral energy distributions could serve as powerful priors for the super-resolution task (\citealt{ion1}). Expanding the training set through future \gls{jwst} programs, particularly those targeting higher spectral resolution ($R\sim2700$) over wider areas, would directly improve reconstruction fidelity. Finally, extending the model to handle absorption-line features and continuum diagnostics would broaden its applicability to the full diversity of galaxy populations observed in current and upcoming surveys.

\section{Acknowledgments}
We thank \textit{Simeon Bird}, \textit{Jeffrey W. Kruk}, \textit{Motahareh Pourrahimi}, \textit{Vladimir Vuskovic}, \textit{Alexander de la Vega}, and the \textit{AstroAI} team at Harvard Center for Astronomy for their valuable comments during the preparation of this manuscript. We are also grateful to the anonymous referee, whose careful report significantly improved this work. This work does not relate to the positions of Hooshang Nayyeri at Amazon.

We thank the \gls{jades} team for making their data publicly available. This work is based on observations made with the NASA/ESA/CSA James Webb Space Telescope. The data were obtained from the Mikulski Archive for Space Telescopes at the Space Telescope Science Institute, which is operated by the Association of Universities for Research in Astronomy, Inc., under NASA contract NAS5-03127 for \gls{jwst}. These observations are associated with programs 1180, 1181, 1210, 1286, 1287, and 3215. This work is also based on observations taken by the \gls{candels} Multi-Cycle Treasury Program with the NASA/ESA HST, which is operated by the Association of Universities for Research in Astronomy, Inc., under NASA contract NAS5-26555.

\appendix

\section{Hyperparameter Optimization} \label{sec:optimization}

The sensitivity of each stage to its free parameters was measured with a hyperparameter search, one sweep per stage, run with the \textit{Weights \& Biases} framework (\citealt{wandb}). All models were implemented in PyTorch and trained on an NVIDIA GeForce RTX 5090 \gls{gpu}. Each samples its space at random and stops unpromising trials early with hyperband, and the sweeps ran in sequence, each reading frozen weights for the stages above it. Together they comprise 122 trials and 153 \gls{gpu}-hours: 40 trials at 61 \gls{gpu}-hours for \gls{sr1}, 42 at 44 for the redshift head, and 40 at 48 for \gls{sr2}.

\begin{figure*}[ht!]
    \centering
    \includegraphics[width=\linewidth]{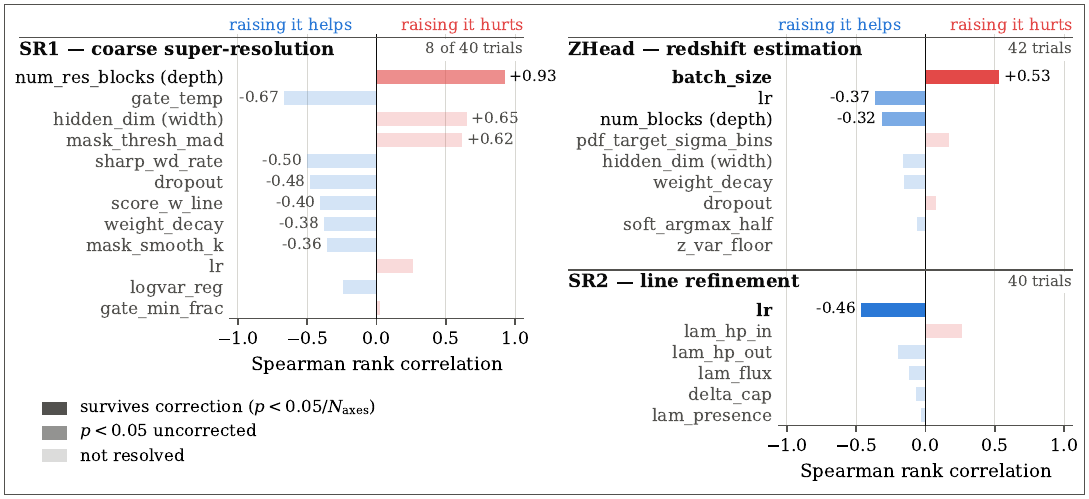}
    \caption{
    Hyperparameter sensitivity of the three pipeline stages. For each searched parameter, the bar gives the Spearman rank correlation with that stage's validation metric, oriented so that lower is better; a positive correlation therefore means that raising the parameter degrades the stage. Shading encodes a two-sided permutation test: solid where the correlation survives a Bonferroni threshold of $0.05/N$ for the $N$ axes of that sweep, half-toned where it clears an uncorrected $0.05$ alone, and faint where the trial count does not resolve it. The redshift and \gls{sr2} groups use every trial, ordered by how far hyper-band allowed each to run and by the final metric within a survival tier; the \gls{sr1} group uses its eight full-length trials, ranked on validation mean squared error because its own selection score contains searched loss weights.}
    
    \label{fig:sweep}
\end{figure*}

Two parameters are resolved at the corrected threshold. The redshift head responds to batch size ($\rho = +0.53$, $p = 6\times10^{-4}$), with smaller batches yielding a lower median $|\Delta z|/(1+z)$, and \gls{sr2} responds to learning rate ($\rho = -0.46$, $p = 2\times10^{-3}$).

On the present evidence the stages are not delicately tuned: the behavior of the pipeline is set considerably more by the preprocessing corrections of \S\ref{sec:preproc} than by any individual hyperparameter within the ranges searched.

\section{Software and data availability} \label{sec:software}
The framework presented here is released as \texttt{specsr}, an open-source Python package (\url{https://github.com/aryana-haghjoo/specsr}), archived at \url{https://doi.org/10.5281/zenodo.21943196} and installable with \texttt{pip install specsr}. Documentation and tutorials are available at \url{https://aryana-haghjoo.github.io/specsr}. The trained weights are hosted on the Hugging Face (\url{https://huggingface.co/aryana-haghjoo/specsr}) and are downloaded on first use. Every result and figure presented in this paper can be reproduced by the package. The \gls{jades} \gls{dr}4 spectra were obtained from \url{https://jades.herts.ac.uk/DR4/}. The \gls{jades} and \gls{candels} science products are available at MAST via \dataset[doi:10.17909/8tdj-8n28]{https://doi.org/10.17909/8tdj-8n28} and \dataset[doi:10.17909/T94S3X]{https://doi.org/10.17909/T94S3X}, respectively.
\bibliography{references}

@article{Curtis-Lake2025,
  title={JADES data release 4--Paper i. Sample selection, observing strategy and redshifts of the complete spectroscopic sample},
  author={Curtis-Lake, Emma and Cameron, Alex J and Bunker, Andrew J and Scholtz, Jan and Carniani, Stefano and Parlanti, Eleonora and D’Eugenio, Francesco and Jakobsen, Peter and Willmer, Christopher NA and Arribas, Santiago and others},
  journal={Monthly Notices of the Royal Astronomical Society},
  volume={549},
  number={4},
  pages={stag836},
  year={2026},
  publisher={Oxford University Press}
}

@article{Darvish2015,
  title={Spectroscopic Study of Star-forming Galaxies in Filaments and the Field at z~ 0.5: Evidence for Environmental Dependence of Electron Density},
  author={Darvish, Behnam and Mobasher, Bahram and Sobral, David and Hemmati, Shoubaneh and Nayyeri, Hooshang and Shivaei, Irene},
  journal={The Astrophysical Journal},
  volume={814},
  number={2},
  pages={84},
  year={2015},
  publisher={The American Astronomical Society}
}

@article{DESI2025,
  title={DESI 2024 VI: cosmological constraints from the measurements of baryon acoustic oscillations},
  author={Adame, Armand G and Aguilar, J and Ahlen, Steven and Alam, Shadab and Alexander, David M and Alvarez, M and Alves, Ot{\'a}vio and Anand, Abhijeet and Andrade, Uendert and Armengaud, Eric and others},
  journal={Journal of Cosmology and Astroparticle Physics},
  volume={2025},
  number={02},
  pages={021},
  year={2025},
  publisher={IOP Publishing}
}

@article{hemmati2019,
  title={Photometric redshift calibration requirements for WFIRST weak-lensing cosmology: predictions from CANDELS},
  author={Hemmati, Shoubaneh and Capak, Peter and Masters, Daniel and Davidzon, Iary and Dor{\`e}, Olivier and Kruk, Jeffrey and Mobasher, Bahram and Rhodes, Jason and Scolnic, Daniel and Stern, Daniel},
  journal={The Astrophysical Journal},
  volume={877},
  number={2},
  pages={117},
  year={2019},
  publisher={The American Astronomical Society}
}

@article{woodrum2025,
  title={JADES: The Star Formation and Dust Attenuation Properties of Galaxies at 3< z< 7},
  author={Woodrum, Charity and Shivaei, Irene and Witstok, Joris and Saxena, Aayush and Simmonds, Charlotte and Scholtz, Jan and Bhatawdekar, Rachana and Bunker, Andrew J and Carniani, St{\'e}fano and Charlot, Stephane and others},
  journal={arXiv preprint arXiv:2510.00235},
  year={2025}
}

@article{Alavi2026,
  title={UV Spectral Slope and Nebular Dust Attenuation in Dwarf Galaxies at 1.4< z< 2.6},
  author={Alavi, Anahita and Siana, Brian and Teplitz, Harry I and Gburek, Timothy and Colbert, James and Mehta, Vihang and Emami, Najmeh and Freeman, William R and Richard, Johan and Kim, Keunho},
  journal={The Astrophysical Journal},
  volume={997},
  number={2},
  pages={290},
  year={2026},
  publisher={The American Astronomical Society}
}

@article{Shivaei2020,
  title={The MOSDEF survey: the variation of the dust attenuation curve with metallicity},
  author={Shivaei, Irene and Reddy, Naveen and Rieke, George and Shapley, Alice and Kriek, Mariska and Battisti, Andrew and Mobasher, Bahram and Sanders, Ryan and Fetherolf, Tara and Azadi, Mojegan and others},
  journal={The Astrophysical Journal},
  volume={899},
  number={2},
  pages={117},
  year={2020},
  publisher={The American Astronomical Society}
}

@article{Shivaei2016,
  title={THE MOSDEF SURVEY: THE STRONG AGREEMENT BETWEEN H $\alpha$ AND UV-TO-FIR STAR FORMATION RATES FOR z~ 2 STAR-FORMING GALAXIES},
  author={Shivaei, Irene and Kriek, Mariska and Reddy, Naveen A and Shapley, Alice E and Barro, Guillermo and Conroy, Charlie and Coil, Alison L and Freeman, William R and Mobasher, Bahram and Siana, Brian and others},
  journal={The Astrophysical Journal Letters},
  volume={820},
  number={2},
  pages={L23},
  year={2016},
  publisher={The American Astronomical Society}
}

@article{khostovan2024,
  title={Evolution of H $\alpha$ equivalent widths from z~ 0.4-2.2: implications for star formation and legacy surveys with Roman and Euclid},
  author={Khostovan, Ali Ahmad and Malhotra, Sangeeta and Rhoads, James E and Sobral, David and Harish, Santosh and Tilvi, Vithal and Coughlin, Alicia and Rezaee, Saeed},
  journal={Monthly Notices of the Royal Astronomical Society},
  volume={535},
  number={4},
  pages={2903--2926},
  year={2024},
  publisher={Oxford University Press}
}

@article{shen2025,
  title={NGDEEP: The Star Formation and Ionization Properties of Galaxies at 1.7< z< 3.4},
  author={Shen, Lu and Papovich, Casey and Matharu, Jasleen and Pirzkal, Nor and Hu, Weida and Berg, Danielle A and Bagley, Micaela B and Backhaus, Bren E and Cleri, Nikko J and Dickinson, Mark and others},
  journal={The Astrophysical Journal Letters},
  volume={980},
  number={2},
  pages={L45},
  year={2025},
  publisher={The American Astronomical Society}
}

@article{sobral2009,
  title={HiZELS: a high-redshift survey of H$\alpha$ emitters--II. The nature of star-forming galaxies at z= 0.84},
  author={Sobral, D and Best, PN and Geach, JE and Smail, Ian and Kurk, J and Cirasuolo, M and Casali, M and Ivison, RJ and Coppin, Kristen and Dalton, GB},
  journal={Monthly Notices of the Royal Astronomical Society},
  volume={398},
  number={1},
  pages={75--90},
  year={2009},
  publisher={Blackwell Publishing Ltd Oxford, UK}
}

@article{wang2019,
  title={ATLAS probe: Breakthrough science of galaxy evolution, cosmology, Milky Way, and the Solar System},
  author={Wang, Yun and Robberto, Massimo and Dickinson, Mark and Hillenbrand, Lynne A and Fraser, Wesley and Behroozi, Peter and Brinchmann, Jarle and Chuang, Chia-Hsun and Cimatti, Andrea and Content, Robert and others},
  journal={Publications of the Astronomical Society of Australia},
  volume={36},
  pages={e015},
  year={2019},
  publisher={Cambridge University Press}
}

@article{faisst2018empirical,
  title={Empirical Modeling of the Redshift Evolution of the [N II]/H $\alpha$ Ratio for Galaxy Redshift Surveys},
  author={Faisst, Andreas L and Masters, Daniel and Wang, Yun and Merson, Alexander and Capak, Peter and Malhotra, Sangeeta and Rhoads, James E},
  journal={The Astrophysical Journal},
  volume={855},
  number={2},
  pages={132},
  year={2018},
  publisher={The American Astronomical Society}
}

@article{agn_misclass,
  title={VLT-MUSE spectroscopy of AGNs misclassified by BPT diagnostic or with weak emission lines},
  author={Agostino, Christopher J and Salim, Samir and Boquien, M{\'e}d{\'e}ric and Janowiecki, Steven and Salas, H{\'e}ctor and Couto, Guillherme S},
  journal={Monthly Notices of the Royal Astronomical Society},
  volume={526},
  number={3},
  pages={4455--4466},
  year={2023},
  publisher={Oxford University Press}
}

@article{scholte2025jwst,
  title={The JWST EXCELS survey: probing strong-line diagnostics and the chemical evolution of galaxies over cosmic time using Te-metallicities},
  author={Scholte, D and Cullen, F and Carnall, AC and Arellano-C{\'o}rdova, KZ and Stanton, TM and Barrufet, L and Begley, R and Bondestam, C and Donnan, CT and Dunlop, JS and others},
  journal={Monthly Notices of the Royal Astronomical Society},
  volume={540},
  number={2},
  pages={1800--1826},
  year={2025},
  publisher={Oxford University Press}
}

@article{laseter2024jades,
  title={JADES: Detecting [OIII] $\lambda$ 4363 emitters and testing strong line calibrations in the high-z Universe with ultra-deep JWST/NIRSpec spectroscopy up to z~ 9.5},
  author={Laseter, Isaac H and Maseda, Michael V and Curti, Mirko and Maiolino, Roberto and D’Eugenio, Francesco and Cameron, Alex J and Looser, Tobias J and Arribas, Santiago and Baker, William M and Bhatawdekar, Rachana and others},
  journal={Astronomy \& Astrophysics},
  volume={681},
  pages={A70},
  year={2024},
  publisher={EDP Sciences}
}

@article{nakajima2023jwst,
  title={JWST census for the mass--metallicity star formation relations at z= 4--10 with self-consistent flux calibration and proper metallicity calibrators},
  author={Nakajima, Kimihiko and Ouchi, Masami and Isobe, Yuki and Harikane, Yuichi and Zhang, Yechi and Ono, Yoshiaki and Umeda, Hiroya and Oguri, Masamune},
  journal={The Astrophysical Journal Supplement Series},
  volume={269},
  number={2},
  pages={33},
  year={2023},
  publisher={The American Astronomical Society}
}

@article{clarke2024star,
  title={The star-forming main sequence in JADES and CEERS at z> 1.4: investigating the burstiness of star formation},
  author={Clarke, Leonardo and Shapley, Alice E and Sanders, Ryan L and Topping, Michael W and Brammer, Gabriel B and Bento, Trinity and Reddy, Naveen A and Kehoe, Emily},
  journal={The Astrophysical Journal},
  volume={977},
  number={1},
  pages={133},
  year={2024},
  publisher={The American Astronomical Society}
}

@article{astroclip,
  title={AstroCLIP: a cross-modal foundation model for galaxies},
  author={Parker, Liam and Lanusse, Francois and Golkar, Siavash and Sarra, Leopoldo and Cranmer, Miles and Bietti, Alberto and Eickenberg, Michael and Krawezik, Geraud and McCabe, Michael and Morel, Rudy and others},
  journal={Monthly Notices of the Royal Astronomical Society},
  volume={531},
  number={4},
  pages={4990--5011},
  year={2024},
  publisher={Oxford University Press}
}

@article{agn_classification,
  title={The host galaxies and classification of active galactic nuclei},
  author={Kewley, Lisa J and Groves, Brent and Kauffmann, Guinevere and Heckman, Tim},
  journal={Monthly Notices of the Royal Astronomical Society},
  volume={372},
  number={3},
  pages={961--976},
  year={2006},
  publisher={Blackwell Publishing Ltd Oxford, UK}
}

@article{HN_blending,
  title={Effects of [N ii] and H $\alpha$ line blending on the WFIRST Galaxy redshift survey},
  author={Martens, Daniel and Fang, Xiao and Troxel, MA and DeRose, Joe and Hirata, Christopher M and Wechsler, Risa H and Wang, Yun},
  journal={Monthly Notices of the Royal Astronomical Society},
  volume={485},
  number={1},
  pages={211--228},
  year={2019},
  publisher={Oxford University Press}
}

@article{highly_related,
  title={Galaxy spectroscopy without spectra: Galaxy properties from photometric images with conditional diffusion models},
  author={Doorenbos, Lars and Sextl, Eva and Heng, Kevin and Cavuoti, Stefano and Brescia, Massimo and Torbaniuk, Olena and Longo, Giuseppe and Sznitman, Raphael and M{\'a}rquez-Neila, Pablo},
  journal={The Astrophysical Journal},
  volume={977},
  number={1},
  pages={131},
  year={2024},
  publisher={The American Astronomical Society}
}

@article{shochandfeedback_1,
  title={Upper boundaries of active galactic nucleus regions in optical diagnostic diagrams},
  author={Ji, Xihan and Yan, Renbin and Riffel, Rog{\'e}rio and Drory, Niv and Zhang, Kai},
  journal={Monthly Notices of the Royal Astronomical Society},
  volume={496},
  number={2},
  pages={1262--1277},
  year={2020},
  publisher={Oxford University Press}
}

@article{shochandfeedback_2,
  title={The MAPPINGS III library of fast radiative shock models},
  author={Allen, Mark G and Groves, Brent A and Dopita, Michael A and Sutherland, Ralph S and Kewley, Lisa J},
  journal={The Astrophysical Journal Supplement Series},
  volume={178},
  number={1},
  pages={20--55},
  year={2008}
}

@article{shochandfeedback_3,
  title={Galaxy-wide shocks in late-merger stage luminous infrared galaxies},
  author={Rich, Jeffrey A and Kewley, Lisa J and Dopita, Michael A},
  journal={The Astrophysical Journal},
  volume={734},
  number={2},
  pages={87},
  year={2011},
  publisher={The American Astronomical Society}
}

@article{kinematics_1,
  title={The SINS survey: SINFONI integral field spectroscopy of z~ 2 star-forming galaxies},
  author={F{\"o}rster Schreiber, NM and Genzel, R and Bouch{\'e}, N and Cresci, G and Davies, R and Buschkamp, P and Shapiro, K and Tacconi, LJ and Hicks, EKS and Genel, S and others},
  journal={The Astrophysical Journal},
  volume={706},
  number={2},
  pages={1364--1428},
  year={2009},
  publisher={The American Astronomical Society}
}

@article{hemmati2022,
  title={Deblending galaxies with generative adversarial networks},
  author={Hemmati, Shoubaneh and Huff, Eric and Nayyeri, Hooshang and Fert{\'e}, Agn{\`e}s and Melchior, Peter and Mobasher, Bahram and Rhodes, Jason and Shahidi, Abtin and Teplitz, Harry},
  journal={The Astrophysical Journal},
  volume={941},
  number={2},
  pages={141},
  year={2022},
  publisher={The American Astronomical Society}
}

@book{electron_density_sii,
  title={Astrophysics of gaseous nebulae and active galactic nuclei, 2nd},
  author={Osterbrock, Donald E and Ferland, Gary J},
  volume={15},
  year={2006},
  publisher={University Science Books, Sausalito, CA}
}

@article{cosmology_1,
  title={Euclid definition study report},
  author={Laureijs, Rene and Amiaux, J{\'e}r{\^o}me and Arduini, S and Augueres, J-L and Brinchmann, J and Cole, R and Cropper, M and Dabin, C and Duvet, L and Ealet, AJAPA and others},
  journal={arXiv preprint arXiv:1110.3193},
  year={2011}
}

@article{jades_overview,
  title={Overview of the JWST advanced deep extragalactic survey (JADES)},
  author={Eisenstein, Daniel J and Willott, Chris and Alberts, Stacey and Arribas, Santiago and Bonaventura, Nina and Bunker, Andrew J and Cameron, Alex J and Carniani, Stefano and Charlot, Stephane and Curtis-Lake, Emma and others},
  journal={The Astrophysical Journal Supplement Series},
  volume={283},
  number={1},
  pages={6},
  year={2026},
  publisher={The American Astronomical Society}
}

@article{jades_nircam,
  title={JADES initial data release for the Hubble Ultra Deep Field: revealing the faint infrared sky with deep JWST NIRCam imaging},
  author={Rieke, Marcia J and Robertson, Brant and Tacchella, Sandro and Hainline, Kevin and Johnson, Benjamin D and Hausen, Ryan and Ji, Zhiyuan and Willmer, Christopher NA and Eisenstein, Daniel J and Pusk{\'a}s, D{\'a}vid and others},
  journal={The Astrophysical Journal Supplement Series},
  volume={269},
  number={1},
  pages={16},
  year={2023},
  publisher={The American Astronomical Society}
}

@article{jades_nircam_2,
  title={The JADES Origins Field: a new JWST deep field in the JADES second NIRCam data release},
  author={Eisenstein, Daniel J and Johnson, Benjamin D and Robertson, Brant and Tacchella, Sandro and Hainline, Kevin and Jakobsen, Peter and Maiolino, Roberto and Bonaventura, Nina and Bunker, Andrew J and Cameron, Alex J and others},
  journal={The Astrophysical Journal Supplement Series},
  volume={281},
  number={2},
  pages={50},
  year={2025},
  publisher={The American Astronomical Society}
}

@article{jades_nirspec,
  title={JADES NIRSpec initial data release for the Hubble Ultra Deep Field: redshifts and line fluxes of distant galaxies from the deepest JWST Cycle 1 NIRSpec multi-object spectroscopy},
  author={Bunker, Andrew J and Cameron, Alex J and Curtis-Lake, Emma and Jakobsen, Peter and Carniani, Stefano and Curti, Mirko and Witstok, Joris and Maiolino, Roberto and D’Eugenio, Francesco and Looser, Tobias J and others},
  journal={Astronomy \& Astrophysics},
  volume={690},
  pages={A288},
  year={2024},
  publisher={EDP Sciences}
}

@article{jades_nirspec_2,
  title={JADES Data Release 3: NIRSpec/microshutter assembly spectroscopy for 4000 galaxies in the GOODS fields},
  author={d’Eugenio, Francesco and Cameron, Alex J and Scholtz, Jan and Carniani, Stefano and Willott, Chris J and Curtis-Lake, Emma and Bunker, Andrew J and Parlanti, Eleonora and Maiolino, Roberto and Willmer, Christopher NA and others},
  journal={The Astrophysical Journal Supplement Series},
  volume={277},
  number={1},
  pages={4},
  year={2025},
  publisher={The American Astronomical Society}
}

@misc{wandb,
  author       = {Biewald, Lukas},
  title        = {Experiment Tracking with Weights and Biases},
  year         = {2020},
  howpublished = {\url{https://www.wandb.com/}}
}

@article{wide_gto,
  title={The NIRSpec wide GTO survey},
  author={Maseda, Michael V and de Graaff, Anna and Franx, Marijn and Rix, Hans-Walter and Carniani, Stefano and Laseter, Isaac and Dudzevi{\v{c}}i{\=u}t{\.e}, Ugn{\.e} and Rawle, Tim and Parlanti, Eleonora and Arribas, Santiago and others},
  journal={Astronomy \& Astrophysics},
  volume={689},
  pages={A73},
  year={2024},
  publisher={EDP Sciences}
}

@article{sogol_1,
  title={The Application of Manifold Learning to a Selection of Different Galaxy Populations and Scaling Relation Analysis},
  author={Sanjaripour, Sogol and Hemmati, Shoubaneh and Mobasher, Bahram and Canalizo, Gabriela and Barish, Barry C and Shivaei, Irene and Coil, Alison L and Chartab, Nima and Jafariyazani, Marziye and Reddy, Naveen A and others},
  journal={The Astrophysical Journal},
  volume={977},
  number={2},
  pages={202},
  year={2024},
  publisher={The American Astronomical Society}
}

@article{shooby_agn,
  title={Reducing the Dimensions of Active Galactic Nuclei Light-curve Manifolds},
  author={Hemmati, Shoubaneh and Krick, Jessica and Stern, Daniel and Desai, Vandana and Faisst, Andreas and Mart{\'\i}n-Garc{\'\i}a, Lucas and Gorjian, Varoujan and Haghjoo, Aryana and Nikakhtar, Farnik and Raen, Troy and others},
  journal={The Astrophysical Journal},
  volume={998},
  number={1},
  pages={130},
  year={2026},
  publisher={The American Astronomical Society}
}

@article{BPT_original,
  title={Classification parameters for the emission-line spectra of extragalactic objects.},
  author={Baldwin, Jack A and Phillips, Mark M and Terlevich, Roberto},
  journal={Publications of the Astronomical Society of the Pacific},
  volume={93},
  number={551},
  pages={5--19},
  year={1981},
  publisher={The Astronomical Society of the Pacific}
}

@article{applicationsAIinAstro,
  title={Applications of AI in Astronomy},
  author={Djorgovski, S George and Mahabal, AA and Graham, Matthew J and Polsterer, Kai and Krone-Martins, Alberto},
  journal={arXiv preprint arXiv:2212.01493},
  year={2022}
}

@article{jwst,
  title={The james webb space telescope},
  author={Gardner, Jonathan P and Mather, John C and Clampin, Mark and Doyon, Rene and Greenhouse, Matthew A and Hammel, Heidi B and Hutchings, John B and Jakobsen, Peter and Lilly, Simon J and Long, Knox S and others},
  journal={Space Science Reviews},
  volume={123},
  number={4},
  pages={485--606},
  year={2006},
  publisher={Springer}
}

@article{SDSS,
  title={Sloan Digital Sky Survey. V. Pioneering Panoptic Spectroscopy},
  author={Kollmeier, Juna A and Rix, Hans-Walter and Aerts, Conny and Aird, James and Vera Alfaro, Pablo and Almeida, Andr{\'e}s and Anderson, Scott F and Arseneau, Stefan M and Assef, Roberto J and Aviram, Shir and others},
  journal={The Astronomical Journal},
  volume={171},
  number={1},
  pages={52},
  year={2026},
  publisher={The American Astronomical Society}
}

@inproceedings{DESI,
  title={Overview of the dark energy spectroscopic instrument},
  author={Martini, Paul and Bailey, Stephen and Besuner, Robert W and Brooks, David and Doel, Peter and Edelstein, Jerry and Eisenstein, Daniel and Flaugher, Brenna and Gutierrez, Gaston and Harris, Stewart E and others},
  booktitle={Ground-based and Airborne Instrumentation for Astronomy VII},
  volume={10702},
  pages={410--420},
  year={2018},
  organization={SPIE}
}

@inproceedings{euclid,
  title={The euclid mission},
  author={Laureijs, Ren{\'e} J and Duvet, Ludovic and Sanz, Isabel Escudero and Gondoin, Philippe and Lumb, David H and Oosterbroek, Tim and Criado, Gonzalo Saavedra},
  booktitle={Space Telescopes and Instrumentation 2010: Optical, Infrared, and Millimeter Wave},
  volume={7731},
  pages={453--458},
  year={2010},
  organization={SPIE}
}

@article{roman_1,
  title={WFIRST-2.4: what every astronomer should know},
  author={Spergel, D and Gehrels, N and Breckinridge, J and Donahue, M and Dressler, A and Gaudi, BS and Greene, T and Guyon, O and Hirata, C and Kalirai, J and others},
  journal={arXiv preprint arXiv:1305.5425},
  year={2013}
}

@article{GOODS,
  title={The Great Observatories Origins Deep Survey: initial results from optical and near-infrared imaging},
  author={Giavalisco, M and Ferguson, HC and Koekemoer, AM and Dickinson, M and Alexander, DM and Bauer, FE and Bergeron, J and Biagetti, C and Brandt, WN and Casertano, S and others},
  journal={The Astrophysical Journal Letters},
  volume={600},
  number={2},
  pages={L93--L98},
  year={2004}
}

@article{ion1,
  title={AION-1: Omnimodal foundation model for astronomical sciences},
  author={Parker, Liam and Lanusse, Francois and Shen, Jeff and Liu, Ollie and Hehir, Tom and Sarra, Leopoldo and Meyer, Lucas and Bowles, Micah and Wagner-Carena, Sebastian and Qu, Helen and others},
  journal={Advances in Neural Information Processing Systems},
  volume={38},
  pages={95386--95428},
  year={2026}
}

@inproceedings{roman_2,
  title={Survey science with the Nancy Grace Roman Space Telescope wide field instrument},
  author={Schlieder, Joshua E and Barclay, Thomas and Barnes, Amethyst and Bray, Evan and Choi, Ami and Cromey, Benjamin and Delker, Thomas and Finch, Timothy and Frater, Eric H and Hill, Robert J and others},
  booktitle={Space Telescopes and Instrumentation 2024: Optical, Infrared, and Millimeter Wave},
  volume={13092},
  pages={196--221},
  year={2024},
  organization={SPIE}
}

@article{kinematics_2,
  title={The MOSDEF Survey: Kinematic and Structural Evolution of Star-forming Galaxies at 1.4≤ z≤ 3.8},
  author={Price, Sedona H and Kriek, Mariska and Barro, Guillermo and Shapley, Alice E and Reddy, Naveen A and Freeman, William R and Coil, Alison L and Shivaei, Irene and Azadi, Mojegan and Groot, Laura de and others},
  journal={The Astrophysical Journal},
  volume={894},
  number={2},
  pages={91},
  year={2020},
  publisher={The American Astronomical Society}
}

@book{haghjoo_thesis_mcgill,
  title={Parameter Estimation of the Global 21cm Signal},
  author={Haghjoo, Aryana},
  year={2023},
  publisher={McGill University (Canada)}
}

@article{PettiniPagel2004,
  title={[O iii]/[N ii] as an abundance indicator at high redshift},
  author={Pettini, Max and Pagel, Bernard EJ},
  journal={Monthly Notices of the Royal Astronomical Society},
  volume={348},
  number={3},
  pages={L59--L63},
  year={2004},
  publisher={Blackwell Science Ltd Oxford, UK}
}

@article{tremonti2004,
  title={The origin of the mass-metallicity relation: insights from 53,000 star-forming galaxies in the sloan digital sky survey},
  author={Tremonti, Christy A and Heckman, Timothy M and Kauffmann, Guinevere and Brinchmann, Jarle and Charlot, St{\'e}phane and White, Simon DM and Seibert, Mark and Peng, Eric W and Schlegel, David J and Uomoto, Alan and others},
  journal={The Astrophysical Journal},
  volume={613},
  number={2},
  pages={898--913},
  year={2004}
}

@article{moons,
  title={MOONS: the new multi-object spectrograph for the VLT},
  author={Cirasuolo, Michele and others},
  journal={arXiv preprint arXiv:2009.00628},
  year={2020}
}

@article{fourmost,
  title={4MOST: Project overview and information for the First Call for Proposals},
  author={De Jong, Roelof S and Agertz, Oscar and Berbel, A Agudo and Aird, James and Alexander, David A and Amarsi, Anish and Anders, Friedrich and Andrae, Rene and Ansarinejad, Behzad and Ansorge, Wolfgang and others},
  journal={arXiv preprint arXiv:1903.02464},
  year={2019}
}

@article{ginolfi2025,
  title={Inferring redshift and galaxy properties via a multi-task neural net with probabilistic outputs: An application to simulated MOONS spectra},
  author={Ginolfi, Michele and Mannucci, Filippo and Belfiore, Francesco and Marconi, Alessandro and Boardman, Nicholas and Pozzetti, Lucia and Bolzonella, Micol and Di Teodoro, Enrico and Cresci, Giovanni and Wild, Vivienne and others},
  journal={Astronomy \& Astrophysics},
  volume={693},
  pages={A73},
  year={2025},
  publisher={EDP Sciences}
}

@article{zhong2025,
  title={Galaxy Spectra Networks (GaSNet)--III. Reconstructive pre-trained network for spectrum reconstruction, redshift estimate, and anomaly detection},
  author={Zhong, Fucheng and Napolitano, Nicola R and Heneka, Caroline and Krogager, Jens-Kristian and Demarco, Ricardo and Bouch{\'e}, Nicolas F and Loveday, Jonathan and Fritz, Alexander and Verdier, Aur{\'e}lien and Roukema, Boudewijn F and others},
  journal={Monthly Notices of the Royal Astronomical Society},
  volume={543},
  number={1},
  pages={691--708},
  year={2025},
  publisher={Oxford University Press}
}

@inproceedings{arad2016sparse,
  title={Sparse recovery of hyperspectral signal from natural RGB images},
  author={Arad, Boaz and Ben-Shahar, Ohad},
  booktitle={European conference on computer vision},
  pages={19--34},
  year={2016},
  organization={Springer}
}

@article{spectral_reconstruction_survey,
  title={A survey on computational spectral reconstruction methods from RGB to hyperspectral imaging},
  author={Zhang, Jingang and Su, Runmu and Fu, Qiang and Ren, Wenqi and Heide, Felix and Nie, Yunfeng},
  journal={Scientific reports},
  volume={12},
  number={1},
  pages={11905},
  year={2022},
  publisher={Nature Publishing Group UK London}
}

@inproceedings{ledig2017srgan,
  title={Photo-realistic single image super-resolution using a generative adversarial network},
  author={Ledig, Christian and Theis, Lucas and Husz{\'a}r, Ferenc and Caballero, Jose and Cunningham, Andrew and Acosta, Alejandro and Aitken, Andrew and Tejani, Alykhan and Totz, Johannes and Wang, Zehan and others},
  booktitle={2017 IEEE conference on computer vision and pattern recognition (CVPR)},
  pages={105--114},
  year={2017},
  organization={IEEE}
}

@inproceedings{blau2018perception,
  title={The perception-distortion tradeoff},
  author={Blau, Yochai and Michaeli, Tomer},
  booktitle={2018 IEEE/CVF Conference on Computer Vision and Pattern Recognition},
  pages={6228--6237},
  year={2018},
  organization={IEEE}
}

@inproceedings{cohen2018hallucinate,
  title={Distribution matching losses can hallucinate features in medical image translation},
  author={Cohen, Joseph Paul and Luck, Margaux and Honari, Sina},
  booktitle={International conference on medical image computing and computer-assisted intervention},
  pages={529--536},
  year={2018},
  organization={Springer}
}

@article{santini2015,
  title={Stellar masses from the CANDELS survey: the GOODS-South and UDS fields},
  author={Santini, P and Ferguson, HC and Fontana, A and Mobasher, B and Barro, G and Castellano, M and Finkelstein, SL and Grazian, A and Hsu, LT and Lee, B and others},
  journal={The Astrophysical Journal},
  volume={801},
  number={2},
  pages={97},
  year={2015},
  publisher={The American Astronomical Society}
}

@article{barro2019,
  title={The CANDELS/SHARDS multiwavelength catalog in GOODS-N: photometry, photometric redshifts, stellar masses, emission-line fluxes, and star formation rates},
  author={Barro, Guillermo and P{\'e}rez-Gonz{\'a}lez, Pablo G and Cava, Antonio and Brammer, Gabriel and Pandya, Viraj and Moral, Carmen Eliche and Esquej, Pilar and Dom{\'\i}nguez-S{\'a}nchez, Helena and Pampliega, Belen Alcalde and Guo, Yicheng and others},
  journal={The Astrophysical Journal Supplement Series},
  volume={243},
  number={2},
  pages={22},
  year={2019},
  publisher={The American Astronomical Society}
}

@article{grogin2011,
  title={CANDELS: the cosmic assembly near-infrared deep extragalactic legacy survey},
  author={Grogin, Norman A and Kocevski, Dale D and Faber, SM and Ferguson, Henry C and Koekemoer, Anton M and Riess, Adam G and Acquaviva, Viviana and Alexander, David M and Almaini, Omar and Ashby, Matthew LN and others},
  journal={The Astrophysical Journal Supplement Series},
  volume={197},
  number={2},
  pages={35},
  year={2011},
  publisher={The American Astronomical Society}
}

@article{koekemoer2011,
  title={CANDELS: the cosmic assembly near-infrared deep extragalactic legacy survey—the Hubble Space Telescope observations, imaging data products, and mosaics},
  author={Koekemoer, Anton M and Faber, SM and Ferguson, Henry C and Grogin, Norman A and Kocevski, Dale D and Koo, David C and Lai, Kamson and Lotz, Jennifer M and Lucas, Ray A and McGrath, Elizabeth J and others},
  journal={The Astrophysical Journal Supplement Series},
  volume={197},
  number={2},
  pages={36},
  year={2011},
  publisher={The American Astronomical Society}
}

@misc{zestimatr,
  author    = {Haghjoo, Aryana and Woodland, Michele and Lau, Marie Wingyee},
  title     = {zestimatr: Spectroscopic redshift estimation with uncertainty quantification},
  year      = {2026},
  version   = {0.1.1},
  publisher = {Zenodo},
  doi       = {10.5281/zenodo.20835578},
  url       = {https://github.com/aryana-haghjoo/zestimatr}
}
\end{document}